\documentclass[aps,aas,pra,reprint,superscriptaddress,showpacs,floatfix,twocolumn,amsmath,amssymb,nofootinbib,longbibliography,tabulary,color]{revtex4-1}

\usepackage{amsmath,amssymb,graphics,setspace}
\usepackage{bm,soul}
\usepackage[pdftex]{graphicx}
\usepackage{multirow}
\usepackage{color}
\usepackage{courier}
\usepackage{subfigure}
\usepackage{xspace}
\usepackage{tabularx,ragged2e,booktabs}
\usepackage{array} %\newcolumntype{?}{!{\vrule width 1pt}}
\usepackage{ulem}

\usepackage[usenames,dvipsnames,svgnames,table]{xcolor}
\usepackage{natbib}
\usepackage{hyperref} %needs to go before hypersetup
\hypersetup{colorlinks=true,linkcolor=blue,filecolor=magenta,urlcolor=cyan}

\usepackage{dcolumn}
\usepackage{tabulary}
\usepackage{siunitx}
\usepackage{longtable}
\usepackage{enumitem}
\newlist{indenteddesc}{description}{1}
\setlist[indenteddesc]{
  leftmargin=0.8in,  
  rightmargin=0in,
  labelindent=0in, 
  labelwidth=0.5in,
  labelsep=0.1in
}

\makeatletter
\renewcommand{\p@subsection}{}
\renewcommand{\p@subsubsection}{}
\makeatother
\newcommand{\bal}{\begin{align}}
\newcommand{\eal}{\end{align}}

\newcommand {\mathsym}[1]{{}}
\newcommand {\unicode}[1]{{}}

\newcommand{\bfig}{\begin{figure}}
\newcommand{\efig}{\end{figure}}

\newcommand{\pnas}{Proc. Nat. Acad. Sci. USA}

\newcommand{\ES}{S}

\newcommand{\Vs}{V_T}

\newcommand{\ybp}{ b^{\prime} } 

\newcommand{\xa}{a}
\newcommand{\yb}{b}

\newcommand{\x}{x}
\newcommand{\y}{y}
\newcommand{\xp}{ x^{\prime} } 
\newcommand{\yp}{ y^{\prime} } 

\newcommand{\z}{z}

\newcommand{\pone}{ P_1 }

\newcommand{\pxone}{ p_1 } 
\newcommand{\pxtwo}{ p_2 } 
\newcommand{\pxthree}{ p_3 }

\newcommand{\Vg}{ V_{G} }

\newcommand{\Mx}{M_1}
\newcommand{\My}{M_2}
\newcommand{\Fx}{F_1}
\newcommand{\Fy}{F_2}
\newcommand{\lam}{\lambda}

\newcommand{\lami}{\lambda_i}

\newcommand{\Xlam}{ A(\x,\lam) }
\newcommand{\Ylam}{ B(\y,\lam)}
\newcommand{\Xlamp}{A(\xp,\lam)}
\newcommand{\Ylamp}{ B(\yp,\lam)}

\newcommand{\Xlami}{ A(\x|\lami) }
\newcommand{\Ylami}{ B(\y|\lami)}
\newcommand{\Xlampi}{A(\xp|\lami)}
\newcommand{\Ylampi}{ B(\yp|\lami)}

\newcommand{\pxyl}{ p(\lam|\x,\y) } 
\newcommand{\pxypl}{ p(\lam|\x,\yp) } 
\newcommand{\pxpyl}{ p(\lam|\xp,\y) } 
\newcommand{\pxpypl}{ p(\lam|\xp,\yp) } 

\newcommand{\pxyli}{ p(\lami|\x,\y) } 
\newcommand{\pxypli}{ p(\lami|\x,\yp) } 
\newcommand{\pxpyli}{ p(\lami|\xp,\y) } 
\newcommand{\pxpypli}{ p(\lami|\xp,\yp) }

\newcommand{\etal}{{\it et al.~\xspace}}

\newcommand{\dlam}{d\lam}

\newcolumntype{?}{!{\vrule width 1pt}}

\def\pli#1#2{p(\lambda_i|#1,#2)}

\def\half{\frac{1}{2}}

\newcommand{\Mtx}{\hat \Mx}
\newcommand{\Mty}{\hat \My}

\def\rmax{{\rm max}}
\def\rmin{{\rm min}}

\newif\ifveryclean \verycleanfalse

\ifveryclean
   \def\sout#1{\ifhmode\unskip\fi}
\fi

\def\two{two}
\def\four{four}

\begin{document}
%------------------------------------------------------------------------------------------------------------------------------------------------

\title{Relaxed Bell Inequalities with Arbitrary Measurement Dependence for Each Observer}

\author{Andrew S. Friedman}
\email{asf@ucsd.edu}
\affiliation{
Center for Astrophysics and Space Sciences, University of California, San Diego, La Jolla, California 92093, USA}

\author{Alan H. Guth}
\email{guth@ctp.mit.edu}
\affiliation{Center for Theoretical Physics and Department of Physics, Massachusetts Institute of Technology, Cambridge, Massachusetts 02139 USA}

\author{Michael J.W. Hall}
\email{michael.hall@griffith.edu.au}
\affiliation{Centre for Quantum Computation and Communication Technology (Australian Research Council), Centre for Quantum Dynamics, Griffith University, Brisbane, Queensland 4111, Australia}
\affiliation{Department of Theoretical Physics, Research School of Physics and Engineering,
Australian National University, Canberra ACT 0200, Australia}

\author{David I. Kaiser}
\email{dikaiser@mit.edu}
\affiliation{Center for Theoretical Physics and Department of Physics, Massachusetts Institute of Technology, Cambridge, Massachusetts 02139 USA}

\author{Jason Gallicchio}
\email{jason@hmc.edu}
\affiliation{Department of Physics, Harvey Mudd College, Claremont, California 91711, USA}

\date{\today}

%------------------------------------------------------------------------------------------------------------------------------------------------

\begin{abstract} 
Bell's inequality was originally derived under the assumption
that experimenters are free to select detector settings
independently of any local ``hidden variables" that might affect
the outcomes of measurements on entangled particles. This
assumption has come to be known as ``measurement independence"
(also referred to as ``freedom of choice" 
or ``settings independence"). For a \two-setting, \two-outcome Bell test, we
derive modified Bell inequalities
that relax measurement independence, 
for either or both 
observers, while remaining locally causal.
We describe the loss of measurement independence
for each observer using the parameters $\Mx$ and $\My$, as
defined by Hall in 2010, and also by a more complete description
that adds two new parameters, which we call $\Mtx$ and $\Mty$,
deriving a modified Bell inequality for each description.
These `relaxed'
inequalities subsume those considered in previous work as special
cases, and quantify how much the assumption of measurement
independence needs to be relaxed in order for a locally causal
model to produce a given violation of the standard
Bell-Clauser-Horne-Shimony-Holt
(Bell-CHSH) inequality. We 
show that both relaxed Bell inequalities are tight
bounds on the CHSH parameter by
constructing
locally causal models that saturate 
them. For any given Bell inequality violation, the new \two-parameter and \four-parameter
models each
require significantly less mutual information between the hidden
variables and measurement settings than previous models.
We conjecture that the new models, with optimal parameters, require the minimum possible mutual information for a given Bell violation.
We further argue that, contrary to various claims in the literature,
relaxing freedom of choice need not imply superdeterminism.
\end{abstract}

\maketitle

%-----------------------------------------------------------------------------------------------------------------------------------------------
\section{Introduction}
\label{sec:abs}
%-----------------------------------------------------------------------------------------------------------------------------------------------

Bell's theorem remains a hallmark achievement of modern physics \cite{bell64,clauser69,bell76,bell85,bell87}. 
Since John S. Bell derived his inequality more than 50 years ago \cite{bell64}, numerous experiments with entangled particles have demonstrated clear violations of Bell's inequality, including several recent, state-of-the-art tests \cite{hensen15,shalm15,giustina15,hensen16,rosenfeld17,handsteiner17,PanSatellite,BigBellTest2,rauch18,li18}, each of them consistent with predictions from quantum mechanics. While lending strong empirical support for quantum theory, these tests more directly imply that at least one eminently reasonable assumption required to derive Bell's theorem must fail to hold in the physical world. These include 
local causality---which stipulates that measurement outcomes at one detector cannot depend on the settings or outcomes at a distant detector---and experimenters' ability to select detector settings freely, independent of any ``hidden variables" that might affect the outcomes of measurements.

If one or more assumptions used to derive Bell's theorem are relaxed, this opens up %logical 
``loopholes" whereby 
local ``hidden variable" models could %potentially 
remain consistent with
%explain 
all previous Bell-violating experiments \cite{brunner14,larsson14b}. 
It is therefore crucial to address as many loopholes as possible in a single test.

Some of the best-known loopholes include the possibility of signaling or communication between the detectors regarding the settings or measurement outcomes on each side of the experiment (the ``locality" loophole \cite{aspect82,weihs98}), and the possibility that 
some unknown mechanism is taking advantage of
detector inefficiency to bias the sample of entangled
particles that are detected
(the ``detection" or ``fair-sampling" loophole \cite{pearle70,eberhard93}). There has been considerable interest in conducting experiments that close %both 
either the locality 
%and 
or detection loopholes \cite{aspect82,weihs98,rowe01,ansmann09,scheidl10,hofmann12,giustina13,christensen13}, culminating in several recent
experimental tests that closed both of these loopholes simultaneously \cite{hensen15,giustina15,shalm15,hensen16,rosenfeld17,li18}. 

In addition, Bell's theorem is derived under the assumption that observers have complete freedom to choose detector settings in an experimental test of Bell's inequality. Relaxing this assumption leads to a third, significant loophole.
The ``measurement-independence" loophole (also known as the ``freedom-of-choice" or ``settings-independence" loophole)
has received the least attention to date, though recent theoretical work indicates that the use of Bell tests to exclude local hidden-variable theories  is most vulnerable to this particular loophole \cite{barrett11,koh12,hall10,hall11,banik12,putz14,putz16}. This paper builds on recent interest in theoretical models that relax the measurement-independence assumption \cite{brans88,kofler06,weinstein09,hall10,hall11,barrett11,banik12,koh12,thinh13,pope13,vervoort13,paul14,putz14,gallicchio14,chaves15,aktas15,putz16,weinstein17,weinstein18}, as well as 
recent experiments that constrain such models \cite{scheidl10,gallicchio14,aktas15,handsteiner17,putz16,BigBellTest2,rauch18,li18}.

Even if nature does not exploit the 
measurement-independence loophole,  
addressing the various assumptions experimentally has significant practical relevance for numerous entanglement-based technologies. These include device-independent quantum key distribution \cite{barrett05,pironio09,vazirani14} along with random-number generation and randomness expansion \cite{pironio10,colbeck12,gallego13,Wu2016,leung2017,liu18a, liu18b,bierhorst18,shen18}. In particular, a malicious adversary with knowledge of an opponent's devices could conceivably undermine a variety of quantum information schemes by exploiting the measurement-independence loophole \cite{kofler06,koh12,yuan15a,yuan15b,lid16,tan16,teng16,hall16}. 

%\andy{Referee 1 recommends citing these works which discuss measurement independence \cite{yuan15a,yuan15b,lid16,tan16,teng16}.}

Physicists have constructed theoretical models that can reproduce the quantum singlet-state predictions for measurements on pairs of entangled particles, while obeying local causality, 
by relaxing the assumption of measurement independence---that is, by partially constraining or predicting observers' selections among choices of detector settings \cite{degorre05,hall10,hall11,banik12,barrett11}. The amount of freedom reduction required to reproduce the quantum singlet-state correlations can be quite small, as little as $\simeq 14\%$ deviation from free choice, corresponding to just $\sim 1/15$ of a bit of mutual information between the detector settings and the relevant hidden variables. By contrast, for locally causal models that retain measurement independence, 100\% of determinism or locality must be given up to reproduce the singlet-state correlations~\cite{branciard08,toner03,hall10,hall11,barrett11}, with either generation of one full bit of indeterminism~\cite{branciard08}  or transmission of one full bit of nonlocal signalling~\cite{toner03} being required. Thus the use of Bell experiments to test quantum mechanics---and, by implication, all known quantum-encryption protocols \cite{koh12}---is particularly susceptible to the measurement-independence loophole.

Whereas previous work has assumed identical relaxation of measurement independence for all parties \cite{hall10,hall11} or 100\% freedom for one observer and some nonzero measurement dependence for the other \cite{banik12,barrett11},
in this paper we develop a more general framework that can accommodate different amounts of freedom for each observer. Our motivation stems, in part, from recent efforts to address the measurement-independence loophole experimentally. Some recent experiments have made clever use of human-generated choices \cite{BigBellTest2}, while others have relied upon real-time astronomical observation of light from distant objects (such as quasars) to determine detector settings \cite{handsteiner17,rauch18,li18}. Although any estimation of possible measurement dependence for either of these techniques would be highly model dependent, it is plausible that they would be susceptible to different amounts of measurement dependence. Future Bell-Clauser-Horne-Shimony-Holt (Bell-CHSH) tests, in which observers select distinct methods for determining settings at their detectors, would then generically fall into the general class we analyze here.

We describe the amount of freedom for each
observer by using the parameters $\Mx$ and $\My$ introduced by
Hall in 2010 \cite{hall10}, and we also introduce a more
complete, \four-parameter description that includes two new
parameters, which we call $\Mtx$ and $\Mty$. We consider
\two-setting, \two-outcome Bell-CHSH
tests, and derive upper bounds on the Bell-CHSH parameter for
models that relax measurement independence but maintain local
causality, for both the \two-parameter and \four-parameter
descriptions.
We further show that 
previous bounds for situations with relaxed measurement independence obtained by Hall in Refs.~\cite{hall10,hall11} and by Banik \etal in Ref.~\cite{banik12} are special cases of our more general result. Moreover, we show that both of our new bounds are tight, by constructing \two-parameter and \four-parameter locally causal models that saturate them.  These new models have near-optimal (and conjectured to be optimal) mutual information properties.

The paper is organized as follows. In Sec.~\ref{sec:intro}, we review the assumptions
required for the derivation of Bell's theorem, 
and in Sec.~\ref{sec:meas_dep}, following Refs.~\cite{hall10,hall11}, we introduce a measure, in terms of parameters $\Mx$ and $\My$, with which to quantify each observer's measurement dependence (and also $M$ for overall measurement dependence). In Sec.~\ref{sec:general}, we derive 
a corresponding \two-parameter relaxed Bell inequality. In Sec.~\ref{sec:model} we  
demonstrate that our 
inequality is tight, by constructing a
local and deterministic model that saturates it. In Sec.~\ref{sec:info}, we 
show that, for a given Bell violation,
our model requires significantly less mutual information between measurement settings and hidden variables than previous models, and conjecture that it is in fact optimal in this regard. In Sec.~\ref{sec:4pbound} we introduce a more complete description of measurement dependence that adds two new parameters, $\Mtx$ and $\Mty$.  We generalize our results to a relaxed \four-parameter Bell inequality, and demonstrate that it is tight by presenting a locally causal \four-parameter model that saturates it. Conclusions are presented in Sec.~\ref{sec:conc}. In Appendices \ref{sec:interpol} and \ref{sec:info1} we present a distinct two-parameter model that interpolates between the models of Refs.~\cite{hall10} and \cite{banik12}. We demonstrate that this interpolating model likewise saturates the upper bound of the two-parameter inequality of Sec.~\ref{sec:general}, though it requires significantly more mutual information between the hidden variables and measurement settings to reproduce the predictions of quantum mechanics than does the model presented in Sec.~\ref{sec:model}. Several steps in the derivation of the \four-parameter Bell inequality of Sec.~\ref{sec:4pbound} are presented in Appendix \ref{sec:4boundproof}, and the construction of our \four-parameter model is described in Appendix~\ref{sec:4pmodel}.

%-----------------------------------------------------------------------------------------------------------------------------------------------
\section{Bell's Theorem Assumptions}
\label{sec:intro}
%-----------------------------------------------------------------------------------------------------------------------------------------------

\allowdisplaybreaks

Bell inequalities place restrictions on the statistical correlations between measurements made by two or more observers, under natural assumptions related to local causality and the selection of measurement settings.  For the typical case of two observers, Alice and Bob, we  denote Alice's measurement setting on a given run as $u$ and Bob's as $v$, and the outcomes of their measurements as $a$ and $b$. The statistical correlations between them are then described by a set of joint probability distributions $\{p(a,b|u,v)\}$. To try to account for the correlations within some hidden-variable model, one parameterizes the joint probability distributions in the form
%%%%%%%
\begin{equation}
p (a, b \vert u, v) = \int d\lambda \> p (a, b \vert u, v, \lambda) \> p (\lambda \vert u, v) ,
\label{pabxy}
\end{equation}
where $\lambda$ is a (possibly multi-component) hidden variable that includes among its components any hidden variables that affect the measurement outcomes. Eq.~(\ref{pabxy}) follows from Bayes' theorem and the definition of conditional probability. Note that this equation relies on no assumptions regarding whether events associated with $\lambda$ occur in the past and/or future of various measurements,
or even whether $\lambda$ represents degrees of freedom associated with specific space-time events at all \cite{wiseman14,Beauregard78,argaman10,pricewharton}.

One may constrain Eq.~(\ref{pabxy}) based on additional assumptions regarding locality, determinism, and measurement independence. These assumptions lead to restrictions on the form that the conditional probabilities $p (a, b, \vert u, v ,\lambda)$ may take \cite{bell64,clauser69,bell76,bell85,wiseman14,hall11,hall16,bell87}. The first assumption concerns {\it local causality}: 
\begin{eqnarray}
p(a,b|u,v,\lam) = p(a|u,\lam) \, p(b|v,\lam) .
\label{locality}
\end{eqnarray}
Eq.~(\ref{locality}) assumes the probabilities factorize such that the measurement outcomes on each side depend only on the detector settings on that side and $\lam$. Eq.~(\ref{locality}) may be derived from the joint assumptions of ``outcome independence" and  ``parameter independence'' \cite{wiseman14}, and is motivated by the theoretical and empirical success of relativity. In an ideal Bell test, each measurement event is space-like separated from the setting choice and outcome on the other side, and hence cannot be influenced by them if relativistic causality is valid.

The assumption of {\it determinism} states that the measurement outcomes $a$, $b \in \{-1,1\}$ are given by deterministic functions $a = A(u,\lam)$ and $b=B(v,\lam)$ of the detector settings and $\lam$. Models with locally causal, deterministic outcomes satisfy
\begin{eqnarray}
p (a \vert u, \lambda) = \delta_{a, A (u, \lambda)} \, , \>\> p (b \vert v, \lambda) = \delta_{b, B (v, \lambda)} \, ,
\label{determinism}
\end{eqnarray}
where $\delta_{a,A(u,\lam)}$ and $\delta_{b,B(v,\lam)}$ are Kronecker delta-functions. Determinism thus requires that the conditional outcome probabilities $p(a|u,\lam)$ and $p(b|v,\lam)$ must be either 0 or 1. As demonstrated in Ref.~\cite{hall11}, any locally causal model that satisfies Eq.~(\ref{locality}) for which the outcome probabilities are stochastic (rather than deterministic) functions of the detector setting and $\lambda$ may be 
written in the form of a deterministic model with the same degree of measurement dependence $M$, where $M$ is defined below, in Eq.~(\ref{hallMa}). Hence we restrict attention here to deterministic locally causal models without loss of generality. 

Lastly, Alice and Bob must select detector settings. The assumption known variously as {\it measurement independence}, {\it settings independence}, or {\it freedom of choice} stipulates that the choice of joint detector settings $(u, v)$ is independent of $\lambda$, which includes in its components all the hidden variables that affect measurement outcomes:
%%%%%%
\begin{equation}
p (u, v \vert \lambda) = p (u, v) ,
\label{settingindep1}
\end{equation}
which is equivalent (via Bayes's theorem) to the expression
%%%%%%%%
\begin{equation}
p (\lambda \vert u,v) = p (\lambda) .
\label{settingindep2}
\end{equation}
Equations (\ref{settingindep1}) and (\ref{settingindep2}) imply that Alice's and Bob's choice of detector settings will not be affected by the value of $\lambda$,  and (conversely) that learning Alice's and Bob's detector settings gives no information about the underlying variable $\lambda$  \cite{brans88,kofler06,scheidl10,hall10,hall11,hall16,barrett11,banik12,koh12,thinh13,pope13,paul14,putz14,gallicchio14,chaves15,putz14,putz16,aktas15,conway06,thooft07}. In particular, if Eqs.~(\ref{settingindep1}) and (\ref{settingindep2}) hold, then no hidden third party with the power to affect measurement outcomes can nudge the selections for $u$ and/or $v$ on a given experimental run, nor gain information about these selections from knowledge or manipulation of $\lambda$. We emphasize that these restrictions on third party influences hold regardless of whether we are considering influences that might be causal, retrocausal \cite{Beauregard78,argaman10,pricewharton}, or represent degrees of freedom that are not associated with specific events in 
space-time \cite{wiseman14}.

%-----------------------------------------------------------------------------------------------------------------------------------------------
\section{Quantifying Measurement Independence}
\label{sec:meas_dep}
%-----------------------------------------------------------------------------------------------------------------------------------------------

In this paper we retain the assumption of local causality (and, without loss of generality, determinism), but relax the assumption of measurement independence. We follow the framework established in Refs.~\cite{hall10,hall11} to quantify the degree of relaxation. In particular, we use the variational distance between probability distributions for different settings, $u$ and $v$. 

To motivate this, note from Eq.~\eqref{settingindep2} that measurement dependence corresponds to dependence of the hidden variable distribution $p(\lambda|u,v)$ on $u$ and/or $v$, Alice and/or Bob's measurement settings.  That is, measurement dependence corresponds to $p(\lambda|u_1,v_1) \neq p(\lambda|u_2,v_2)$ for at least some choice of $u_1,u_2,v_1,v_2$. A well-known way to quantify the difference between two probability distributions $p(\lambda)$ and $q(\lambda)$ is via the variational or trace distance \cite{nielsen_and_chuang_book,FuchsVanDeGraaf}, which can be defined as
\begin{equation}
D(p,q) \equiv \int d\lambda\, | p(\lambda) - q(\lambda) | .
\end{equation}
%Further, 
This distance has a simple operational interpretation in terms of an experiment in which one is given a single sample $\lambda$ drawn with equal probability from either the distribution $p$ or the distribution $q$, and then asked which probability distribution was used.  The probability that one can successfully identify the probability distribution, before knowing the value of $\lambda$ that was drawn, is given by \cite{nielsen_and_chuang_book,FuchsVanDeGraaf}
\begin{equation}
P_{\rm distinguish} = \frac{1}{2} \left[1+\frac{1}{2} D(p,q)\right]\, . 
\end{equation}
%(See for example Ref.~\cite{nielsen_and_chuang_book}, which uses the phrase ``trace distance'' for the variational distance.)  
Thus, measurement dependence corresponds to a non-zero distance between $p(\lambda|u_1,v_1)$ and $p(\lambda|u_2,v_2)$ for at least some settings $u_1,u_2, v_1,v_2$, or, equivalently, to a better than 50:50 chance of distinguishing between the measurement settings $(u_1,v_1)$ and $(u_2,v_2)$ on the basis of learning the value of $\lambda$.

We assume that Alice may select her settings from some set $U$, and Bob from some set $V$. Then we may define the overall degree of measurement dependence by 
%%%%%%
\begin{equation}
M \equiv \sup_{u_1,u_2\in U, v_1,v_2\in V} \Bigg\{ \int d \lambda \big\vert p (\lambda \vert u_1, v_1) - p (\lambda \vert u_2, v_2) \big \vert \Bigg\} \, .
\label{hallMa}
\end{equation}
It follows that $M$ quantifies the dependence of the hidden variable distribution on the measurement setting via the maximum distance that can be achieved by varying the settings.  Further, $\half \left(1+\frac{1}{2}M\right)$ determines the maximum probability for distinguishing between pairs of measurement settings. For example, if $M=0$ then there is no measurement dependence: $p(\lambda|u_1,v_1)=p(\lambda|u_2,v_2)$ for all settings $(u_1,v_1)$, $(u_2,v_2)$, and the probability of distinguishing one settings pair from another, based on a sample of $\lambda$, is never better than $\half$.  Thus, the hidden variable contains zero information about the measurement settings.  Conversely, if $M=2$, then there are measurement settings $(u_1,v_1)$ and $(u_2,v_2)$ which can be distinguished with probability one, corresponding to a maximum degree of measurement dependence.

More generally, note that $0 \leq M \leq 2$. 
Measurement independence, with $p (\lambda \vert u, v) = p (\lambda)$ for all $u,v$, yields $M = 0$. The maximum violation of 
measurement independence, $M = 2$, corresponds to the case in which two normalized probability distributions $p (\lambda \vert u_1, v_1)$ and $p (\lambda \vert u_2 , v_2)$ have no overlapping support for any value of $\lambda$. In that case, for each $\lambda$, at most one of the pairs of joint settings $(u_1 , v_1)$ and $(u_2 , v_2)$ may be selected. This implies that if the observers have decided to consider only the two possibilities of joint settings $(u_1 , v_1)$ or $(u_2 , v_2)$, then their choice will be completely dictated by the value of $\lambda$, leaving them no freedom at all.
It is therefore natural to define a corresponding overall degree of freedom of choice $F$ by~\cite{hall10}
\begin{equation} 
F \equiv 1 - \frac{M}{2}.
\label{freedom}
\end{equation}

We may similarly define %\sout{individual} 
one-sided degrees of measurement 
dependence, $\Mx$ and $\My$ \cite{hall10}: 
%%%%%%%%%
\begin{align}
\Mx &\equiv  \sup_{u_1,u_2\in U , v\in V} \Bigg\{ \int \dlam \big| p(\lam|u_1,v) - p(\lam|u_2, v) \big| \Bigg\} \label{hallMa1} \, , \\ 
\My &\equiv \sup_{u\in U, v_1,v_2\in V} \Bigg\{ \int \dlam \big| p(\lam|u,v_1) - p(\lam|u_,v_2) \big| \Bigg\} \label{hallMa2} \, .
\end{align}
Similarly to the case of the overall measurement dependence $M$, discussed above, the one-sided measure $M_1$ quantifies the degree of measurement dependence corresponding to variation of Alice's settings, but with Bob's setting held fixed. Thus, for example, a maximum value $M_1=2$ implies there are measurement settings $(u_1,v)$ and $(u_2,v)$, differing only in Alice's local setting, which can be distinguished by a (hypothetical) measurement of $\lambda$ with probability one.  A similar interpretation holds for $M_2$.

Like $M$, the %\sout{1-sided} 
one-sided parameters are bounded by $0 \le \Mx, \My \le 2$; the corresponding degrees of individual freedom of choice are given by  $\Fx\equiv 1-\Mx/2$ and $\Fy\equiv 1-\My/2$.
The $M$ quantities obey the inequality chain~\cite{hall10} 
%%%%%%
\begin{equation}
{\rm max} \{ \Mx, \My \} \leq M \leq {\rm min} \{ \Mx + \My , 2 \} .
\label{Mineq}
\end{equation}

For experiments in which Alice and Bob each select among two setting choices, 
we may write $u \in \{ x, x' \}$ and $v \in \{ y, y' \}$, and the expressions for $M, \Mx$, and $\My$ simplify to
\begin{eqnarray}
\Mx  =  \max \Bigg\{ & \int  \dlam &\big| p(\lam|\x,\y) - p(\lam|\xp,\y) \big|\ , \nonumber \\
      &  \int  \dlam & \big| p(\lam|\x,\yp) - p(\lam|\xp,\yp) \big| \Bigg\} \, , \label{hallMaCHSH1} \\
\My  =  \max \Bigg\{ & \int  \dlam & \big| p(\lam|\x,\y) - p(\lam|\x,\yp) \big|\ ,\nonumber  \\ 
      &  \int  \dlam & \big| p(\lam|\xp,\y) - p(\lam|\xp,\yp) \big| \Bigg\} \, , \label{hallMaCHSH2}\\
M    =  \max \Bigg\{ &\Mx, & \My,  \int  \dlam \big| p(\lam|\x,\y) - p(\lam|\xp,\yp) \big|\ , \nonumber \\ 
      &  \int  \dlam & \big| p(\lam|\x,\yp) - p(\lam|\xp,\y) \big| \Bigg\} \, . \label{hallMaCHSH} 
\end{eqnarray}
These expressions are useful for calculating the degrees of measurement dependence for the CHSH scenario in later sections. 
%\mg 
We will also consider an alternative measure of correlation, the mutual information between the detector settings and $\lambda$~\cite{hall11,barrett11}, in Sec.~\ref{sec:info}, and  two further parameters related to $M_1$ and $M_2$  in Sec.~\ref{sec:4pbound}.

%-----------------------------------------------------------------------------------------------------------------------------------------------
\section{Relaxed Bell-CHSH Inequality}
\label{sec:general}
%-----------------------------------------------------------------------------------------------------------------------------------------------

In the 
CHSH correlation scenario, Alice and Bob each have two possible measurement settings, $u\in\{x,x'\}$ and $v\in \{y,y'\}$ respectively, each with two corresponding measurement outcomes, $a, b \in \{ -1, 1\}$ respectively.
Defining the correlation function 
%%%%%%%%
\begin{equation}
\langle a b \rangle_{uv} = \sum_{a,b = \pm 1} a b \, p (a, b \vert u, v) ,
\label{abexpuv}
\end{equation}
the CHSH correlation parameter is given by the linear combination 
\cite{clauser69}
%%%%%%%
\begin{eqnarray}
 \ES = \big| \langle \xa \yb \rangle_{xy} + \langle \xa \yb \rangle_{xy'}+ \langle \xa \yb \rangle_{x'y} - \langle \xa \yb \rangle_{x'y'} \big|\ .
\label{bellE}
\end{eqnarray}
Noting that each expectation value can be at most $\pm 1$, the maximum possible value for $\ES$ is 4. However, under the assumptions of local causality (Eq.~(\ref{locality})) and measurement independence (Eq.~(\ref{settingindep2})), one finds the Bell-CHSH inequality~\cite{clauser69}:
\begin{eqnarray}
\ES \le 2 .
\label{bellCHSH}
\end{eqnarray}
By contrast, quantum mechanics predicts a maximum value $\ES_{QM} = 2 \sqrt{2}$ (known as the ``Tsirelson bound" \cite{cirelson80}) for certain choices of detector settings. Therefore quantum mechanics is incompatible with the conjunction of local causality and measurement independence. 

Experiments now routinely measure $\ES > 2$ to high statistical significance, in clear violation of the Bell-CHSH inequality \cite{scheidl10,hensen15,hensen16,shalm15,giustina15,aktas15,rosenfeld17,handsteiner17,PanSatellite,BigBellTest2,rauch18,li18}. The experimental correlations are compatible with quantum predictions. However, alternative models, distinct from quantum mechanics, can also explain the experimental results if one or more of the assumptions leading to Eq.~(\ref{bellCHSH}) fail to hold.

Here we construct a relaxed Bell-CHSH inequality for models that satisfy both local causality and determinism, but relax the  
assumption of measurement independence for each observer. 
For such models, Eqs.~(\ref{pabxy})--(\ref{determinism}) hold but Eqs.~(\ref{settingindep1})--(\ref{settingindep2}) do not. 
The correlation function of Eq.~(\ref{abexpuv}) then takes the form 
%%%%%%%%%%
\begin{equation}
\langle a b \rangle_{uv} = \int d\lambda \, p (\lambda \vert u,v) \> A(u, \lambda) \, B(v, \lambda) \, .
\label{abexpuvM}
\end{equation}
We parameterize the upper bound for the relaxed CHSH-Bell inequality as
\begin{eqnarray}
\ES  & \le & 2 + V ,
\label{bound0}
\end{eqnarray}
where the amount of Bell violation, $V$, will depend on the degree to which measurement independence has been relaxed for Alice and/or Bob. The Tsirelson bound for quantum mechanics, $S_{QM} = 2 \sqrt{2}$, corresponds to a violation 
%%%%%%%%%%%%%%%
\begin{equation}
\Vs = 2 ( \sqrt{2} - 1 ) \simeq 0.828 .
\label{Vsdef}
\end{equation}
We may therefore quantify how much experimental freedom Alice and/or Bob must forfeit in 
locally causal models in order to match the Tsirelson bound, with $V = \Vs$.

For models that obey local causality but relax the overall degree of measurement independence $M$ in Eq.~(\ref{hallMa}), Hall derived the relaxed Bell-CHSH inequality \cite{hall10,hall11}
%%%%%%%%%%
\begin{equation}
S \leq 2 + {\rm min} \big\{ 3 M , 2 \big\}, 
\label{HallS}
\end{equation}
and %gave 
constructed models saturating this bound with $M=\Mx=\My$. Such symmetric models reproduce the Tsirelson bound for quantum mechanics if $\Mx = \My = M= \Vs / 3 \simeq 0.276$, corresponding to degrees of experimental freedom $F=\Fx=\Fy\simeq 86.2\%$, i.e., to Alice and Bob each losing $\simeq 13.8\%$ experimental freedom. Note that neither observer needs to forfeit $100\%$ freedom in order to reach the Tsirelson bound. 

Subsequently, Banik {\it et al.} considered one-sided models in which one observer's freedom is partially reduced while the other observer retains complete freedom: either
$\Mx=M \neq 0$ and $\My = 0$ 
or vice versa \cite{banik12}. Without loss of generality, we may consider $\My=0$. (The converse case $\Mx=0$ follows upon switching observer labels for Alice and Bob, $1 \leftrightarrow 2$.)  Then, the relaxed Bell-CHSH inequality
\begin{equation}
S \leq 2 + \Mx 
\label{BanikS}
\end{equation}
follows, and is saturated by suitable models with $M=\Mx$ and $\My=0$~\cite{banik12}. 
Such 1-sided models reproduce the quantum-mechanical Tsirelson bound with $\Mx = M = \Vs \simeq 0.828$ and $\My = 0$ (or vice versa), corresponding to one observer losing $\Mx / 2 = M/2 \simeq 41.4\%$ freedom. Though such one-sided scenarios require one of the observers to forfeit three times more experimental freedom than in Hall's symmetric case, such models 
similarly 
require considerably less than $\Mx=M = 2$ or $100\%$ reduction of freedom in order to reach the Tsirelson bound.

\begin{figure}
\centering
\begin{tabular}{@{}c@{}c@{}c@{}}
\includegraphics[width=3.2in]{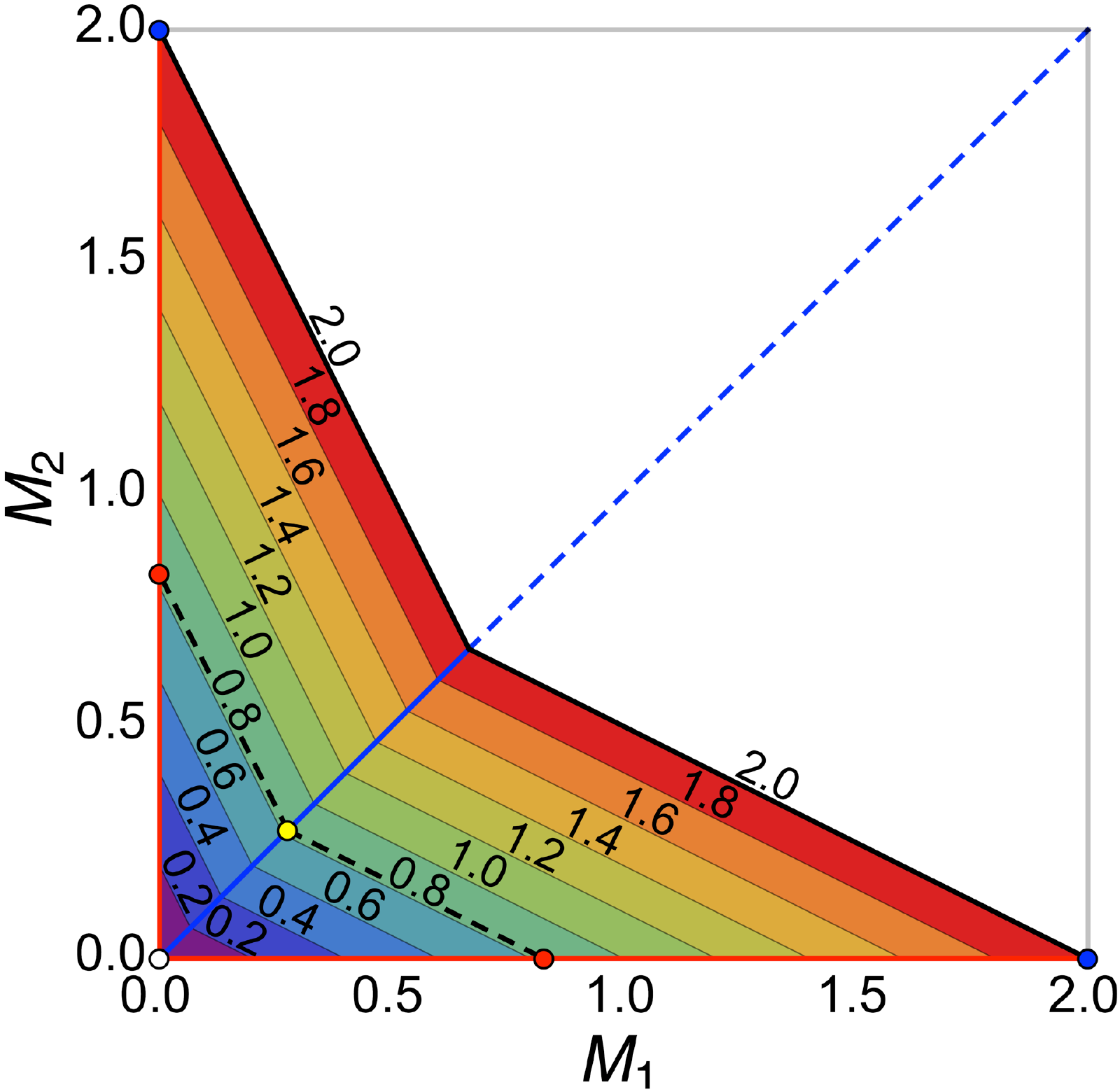}
\end{tabular}
\caption{
This ``freedom square" indicates the minimum degrees of measurement dependence $\Mx$, $\My$ $\in [0,2]$ required for a locally causal model to predict a given violation $V$ of the Bell-CHSH inequality, as per Eq.~(\ref{bound1}). Within the region of principal physical interest, with $V=  \Mx + \My + {\rm min} \{ \Mx, \My \} \leq 2$, 
contours label the amount of CHSH violation $0 \leq V \leq 2$.
(For $\Mx + \My + \min \{ \Mx, \My \} > 2$, i.e., the blank region, Eq.~(\ref{bound1}) yields $V = 2$, corresponding to $S = 4$.) Values of $\Mx$ and $\My$ that yield the Tsirelson bound, with $\Vs = 2 (\sqrt{2} - 1)$, are marked with black dashed lines. 
The solid black lines mark the boundary of the region that yields maximal CHSH violation, $V = 2$. Symmetric models, like those analyzed in Refs.~\cite{hall10,hall11}, lie along the blue (solid and dashed) diagonal line, with $\Mx = \My$ (including the light yellow circle at ($\Mx,\My$)=($\Vs/3,\Vs/3$)), while one-sided models, with $M = \Mx$ and $\My = 0$ or vice versa, as studied in Ref.~\cite{banik12}, lie along the $\Mx$ and $\My$ axes (including the dark red circles at ($\Mx,\My$)=($0,\Vs$) and ($\Vs,0$)). The original Bell-CHSH inequality corresponds to $V = 0$, and is marked by the white
circle at $\Mx = \My = 0$. 
}
\label{FreedomSquarefig}
\end{figure}

In this section we derive a general 
upper bound on $S$ for models that relax measurement independence, as described by the parameters $\Mx,\My \in [0,2]$, while maintaining local causality.  The general two-parameter bound
may be written in the form
%%%%%%%%%
\begin{equation}
S \leq 2 + V_G (\Mx, \My) ,
\label{GeneralS}
\end{equation}
with
\begin{equation}
\Vg(\Mx,\My)
=\min\big\{\Mx + \My + \min\{ \Mx,\My\} ,2\big\} \,.
\label{bound1}
\end{equation}
(In Sec.~\ref{sec:4pbound} we will define two new parameters related to measurement independence, and will describe a four-parameter bound that generalizes Eqs.~\eqref{GeneralS} and \eqref{bound1}.) The bound $V_G$ includes the scenarios studied by Hall in Refs.~\cite{hall10,hall11} and by Banik {\it et al.} in Ref.~\cite{banik12} as special cases. In particular, Eqs.~(\ref{GeneralS})--(\ref{bound1}) reduce to Eq.~(\ref{HallS}) for the case $M=\Mx = \My$ (Hall), and to Eq.~(\ref{BanikS}) for the case $\My=0$ (Banik {\it et al.}). For the general case, we may visualize the amount of 
measurement dependence required of each observer in order to reproduce the Tsirelson bound of quantum mechanics ($V = \Vs$), or the maximal CHSH violation ($V = 2$), as in Fig.~\ref{FreedomSquarefig}.

%%%%%%%%%%%%%%%%%%%%%%%%%%%%%%%%%%%%%%%%%%%%%%%
%%%%%%%%%%%%%%%%%%%%%%%%%%%%%%%%%%%%%%%%%%%%%%%

To derive Eqs.~(\ref{GeneralS})--(\ref{bound1}), we first recall that we may assume the model is deterministic as per Eq.~(\ref{determinism}) without loss of generality. Using Eq.~(\ref{abexpuvM}) to rewrite Eq.~(\ref{bellE}) for the CHSH parameter $S$ then gives
\begin{equation}
\begin{split}
\ES = \Big\vert \int d\lambda \Big[ & \Xlam \Ylam \, \pxyl \\
&\quad  + \Xlamp \Ylam \, \pxpyl \\
&\quad + \Xlam \Ylamp \, \pxypl \\
&\quad - \Xlamp \Ylamp \, \pxpypl \Big] \Big\vert \, .
\end{split}
\label{bellE1}
\end{equation}
We next use a ``plus zero" trick to rewrite Eq.~(\ref{bellE1}) by adding and subtracting identical terms:
\begin{equation}
\begin{split}
\ES  =  \Bigg| \int d\lam \Big\{ & \pxyl  \Big[ \Xlam \Ylam  + \Xlam \Ylamp \Big]   \\
  + &  \pxpyl \Big[ \Xlamp \Ylam - \Xlamp \Ylamp \Big]  \\
  + & \Xlam \Ylamp \Big[ \pxypl - \pxyl \Big]  \\
  - & \Xlamp \Ylamp \Big[ \pxpypl - \pxpyl \Big] \Big\} \Bigg| .
 \end{split}
\label{bellE2}
\end{equation}
Upon using the triangle inequality, we conclude that
\begin{eqnarray}
\ES  \leq T_1 + T_2 + T_3,
\label{bellE3}
\end{eqnarray}
with $T_1$, $T_2$, and $T_3$ given by
\begin{equation}
\begin{split}
T_1 &= \int d\lam \Big|  \pxyl  \Big[ \Xlam \Ylam  + \Xlam \Ylamp \Big]    \\
  &\quad\quad + \pxpyl \Big[ \Xlamp \Ylam - \Xlamp \Ylamp \Big] \Big|,
\end{split}
\label{bellT1}
\end{equation}
\begin{equation}
T_2 = \int d\lam \Big| \Xlam \Ylamp \Big[ \pxypl - \pxyl \Big] \Big|,
\label{bellT2}
\end{equation}
and
\begin{equation}
T_3 =  \int d\lam \Big| \Xlamp \Ylamp \Big[ \pxpypl - \pxpyl \Big] \Big|.
\label{bellT3}
\end{equation}
Since the deterministic outcome functions always have magnitude 
$|\Xlam|=|\Ylamp|=1$, 
$T_2$ in Eq.~(\ref{bellT2}) may be simplified:
\begin{eqnarray}
T_2 = \int d\lam \Big| \pxypl - \pxyl  \Big| \le \My,
\label{bellT2s}
\end{eqnarray}
upon using Eq.~(\ref{hallMaCHSH2}) for $\My$. Similarly,
\begin{eqnarray}
T_3 = \int d\lam \Big| \pxpypl - \pxpyl  \Big| \le \My.
\label{bellT3s}
\end{eqnarray}
Next, Eq.~(\ref{bellT1}) for $T_1$ may be rearranged:
\begin{eqnarray}
T_1 &=& \int d\lam \Bigg|  \Ylam \Big[ \Xlam \pxyl  + \Xlamp \pxpyl \Big]   \nonumber \\
  & +& \Ylamp \Big[ \Xlam \pxyl - \Xlamp \pxpyl \Big] \Bigg|.
\label{bellT1s}
\end{eqnarray}
Again using the triangle inequality and the fact that $|\Xlam|=|\Xlamp|=|\Ylam|=|\Ylamp|=1$ yields
\begin{equation}
\begin{split}
T_1 & \leq  \int d\lam \Bigg\{ \Bigg|  \Ylam \Xlam \Big[ \pxyl  + \frac{\Xlamp}{\Xlam} \pxpyl \Big] \Bigg|   \\
  &\quad\quad  + \Bigg| \Ylamp \Xlam \Big[ \pxyl - \frac{\Xlamp}{\Xlam} \pxpyl \Big] \Bigg| \Bigg\}  \\
  & \leq  \int d\lam \Bigg\{ \Bigg| \pxyl  + \frac{\Xlamp}{\Xlam} \pxpyl  \Bigg|  \\
  &\quad\quad\quad\quad +  \Bigg| \pxyl - \frac{\Xlamp}{\Xlam} \pxpyl \Bigg|   \Bigg\} \, . 
  \end{split}
\label{bellT1s1}
\end{equation}
The quantity $\Xlamp/\Xlam$ is always equal to $+1$ or $-1$ for any value of $\lam$. For either choice, one of the absolute-value arguments in Eq.~(\ref{bellT1s1}) will be $\pxyl + \pxpyl$, and the other will be $\pxyl - \pxpyl$. Thus Eq.~(\ref{bellT1s1}) simplifies to
\begin{eqnarray}
T_1   & \leq  & \int d\lam  \Big\{ \Big| \pxyl  + \pxpyl  \Big| \\
&&\quad \quad + \Big| \pxyl -  \pxpyl \Big|   \Big\}.
\label{bellT1s3}
\end{eqnarray}
But
\begin{eqnarray}
\int d\lam \Big| \pxyl  + \pxpyl  \Big| =2, 
\label{bellT1s4}
\end{eqnarray}
since the function $\pxyl$ is a normalized conditional probability distribution, and 
\begin{eqnarray}
\int d\lam \Big| \pxyl  - \pxpyl  \Big| \leq \Mx, 
\label{bellT1s5}
\end{eqnarray}
upon using Eq.~(\ref{hallMaCHSH1}). Therefore
\begin{eqnarray}
T_1  & \leq & 2 + \Mx.
\label{bellT1s5a}
\end{eqnarray}
Combining Eqs.~(\ref{bellE3}), (\ref{bellT2s}), (\ref{bellT3s}), and Eq.~(\ref{bellT1s5a}), we find
\begin{eqnarray}
\ES & \leq & 2 + \Mx + 2\My .
\label{bellT1s6}
\end{eqnarray}
However, since the formalism makes no distinction between the first and second observer's detectors, we can carry out a parallel set of manipulations, reversing the treatment of $\x$ and $\y$, to similarly obtain
\begin{eqnarray}
\ES & \leq & 2 + \My + 2\Mx .
\label{bellT1s7}
\end{eqnarray}
 Finally, since $\ES$ is less than or equal to the right-hand sides of Eqs.~(\ref{bellT1s6})-(\ref{bellT1s7}), then it must be upper bounded by the minimum of the two, i.e.,
\begin{eqnarray}
\ES & \leq & 2 + \Mx + \My + \min\{ \Mx, \My \}.
\label{bellT1s8}
\end{eqnarray}
Noting that $\ES \le 4$ from Eq.~(\ref{bellE}), we arrive at
\begin{eqnarray}
\ES & \leq & 2 + \min\big\{\Mx + \My + \min\{ \Mx,\My\} ,2\big\}, %\nonumber \\
\label{bellT1s9}
\end{eqnarray}
which is equivalent to Eqs.~(\ref{GeneralS})--(\ref{bound1}), as desired.

%-----------------------------------------------------------------------------------------------------------------------------------------------
\section{Tightness of the General Two-Parameter Bound}
\label{sec:model}
%-----------------------------------------------------------------------------------------------------------------------------------------------

In this section we demonstrate that Eqs.~(\ref{GeneralS})--(\ref{bound1}) yield a tight upper bound on the CHSH parameter $S$ for hidden-variable models that obey local causality while relaxing measurement independence, as described by the parameters $\Mx$ and $\My$. To 
do so, it suffices to show that, for each value of $\Mx$ and $\My$, at least one model exists that saturates $S = 2 + V_G (\Mx, \My)$, with $V_G$ given by Eq.~(\ref{bound1}). 
Hence, similarly to the approach in Refs.~\cite{hall10,hall11,banik12}, we will construct model tables 
with values for Alice's and Bob's measurement outcomes, $A (x , \lambda_i)$ and $B (y , \lambda_i)$, and  conditional probabilities for various values of the hidden variable, $p (\lambda_i \vert x, y)$, subject to the constraint that the $p (\lambda_i \vert x, y)$ are non-negative and properly normalized. We will nonetheless show in Sec.~\ref{sec:4pbound} that if we have additional information about a model, in the form of two new parameters, then we can derive a more general four-parameter bound that can sometimes be tighter than the two-parameter bound of Eqs.~(\ref{GeneralS})--(\ref{bound1}).

In particular, we consider a model with 
a hidden variable $\lambda$ that can take on any of 4 discrete
values, $\lambda_1, \lambda_2, \lambda_3, \lambda_4$, as per 
Tables~\ref{tab:general1a} and~\ref{tab:general2a}.
For this model the deterministic measurement-outcome functions $A (u , \lambda_i)$ and $B (v , \lambda_i)$, for Alice and Bob, respectively, are of the forms defined in Table~\ref{tab:general1a}, where the arbitrary constants $c, d, e, f$ may be any values in  $\{-1, 1\}$.  The conditional probabilities are parameterised by three numbers $p_1$, $p_2$ and $p_3$ as per Table~\ref{tab:general2a}, that can be set to allow different amounts of Bell violation $V_G(\Mx,\My)$ via Eq.~\eqref{bound1} consistent with $S=2+V_G(\Mx,\My)$ from Eq.~\eqref{GeneralS}, for different values of $\Mx$, $\My$. 

The correlations between Alice and Bob's outcomes can be determined from Tables~\ref{tab:general1a} and~\ref{tab:general2a} via Eq.~(\ref{abexpuvM}), and 
we find the CHSH parameter $S$ of Eq.~(\ref{bellE}) takes the form
%%%%%%%
\begin{equation}
S = 2 + 2 \pxone + 4 \pxtwo - 4 \pxthree\, .
\label{Stable0}
\end{equation}
Provided that $\pxone \ge \pxtwo \ge \pxthree$, the degrees of measurement dependence $\Mx$ and $\My$  follow via Eqs.~(\ref{hallMaCHSH1}) and (\ref{hallMaCHSH2}) as 
\begin{eqnarray}
  \Mx &=& {\rm max}\{ 2\pxone, 2\pxone \} = 2 \pxone \ ,\\
  \My &=& {\rm max}\{ 2\pxtwo, 2\pxtwo \} = 2 \pxtwo \ .
  \label{Mtabs12a}
\end{eqnarray}
(For this model, we also find $M = {\rm max}\{\Mx,\My\} = \Mx$.)

%%%%%%%%%%%%%%%%%%%%%%%%%%%%%%%%%%%%%%%%
%TABLE I
%%%%%%%%%%%%%%%%%%%%%%%%%%%%%%%%%%%%%%%%
\linespread{1.0}
\begin{table}[ht]
\centering
\begin{tabular}{| c | c| c| c| c | } 
\toprule[1.5pt]
$\lam_i$       & $\Xlam$ & $\Xlamp$ & $\Ylam$ & $\Ylamp$ \\
\toprule[1.5pt]
${\lam}_1$ & $c$ & $c$ & $c$ & $c$ \\
${\lam}_2$ & $d$ & $-d$ & $d$ & $d$ \\
${\lam}_3$ & $e$ & $e$ & $e$ & $-e$ \\
${\lam}_4$ & $f$ & $-f$ & $-f$ & $f$ \\
\bottomrule[1.25pt]
\end{tabular}\par
\caption{
Deterministic measurement-outcome functions $A (u , \lambda_i)$ and $B (v , \lambda_i)$ for Alice's and Bob's measurements, given $\lambda_i$ with $i = 1, ... , 4$. The values of the measurement outcomes ($c, d, e, f$) are selected  arbitrarily from $\{ -1, 1 \}$.
\label{tab:general1a}
}
\end{table}
%%%%%%%%%%%%%%%%%%%%%%%%%%%%%%%%%%%%%%%%

\def\A{ \frac{1 - \pxone + 2 \pxtwo}{4}}
\def\B{ \frac{1 + \pxone}{4}}
\def\C{\B}
\def\D{\frac{1-\pxone - 2 \pxtwo}{4}}

\def\AP{ \frac{1 {\text -} \pxone + 2 (\pxtwo {\text -}\pxthree)}{4}}
\def\AM{ \frac{1 {\text -} \pxone + 2 (\pxtwo {\text -}\pxthree)}{4}}
\def\BP{ \frac{1 + \pxone + 2\pxthree}{4}}
\def\BM{ \frac{1 + \pxone {\text -} 2\pxthree}{4}}
\def\CP{\BM}
\def\CM{\BP}
\def\DP{ \frac{1{\text -}\pxone {\text -} 2 (\pxtwo{\text -}\pxthree) }{4}}
\def\DM{ \frac{1{\text -}\pxone {\text -} 2 (\pxtwo{\text -}\pxthree) }{4}}

%%%%%%%%%%%%%%%%%%%%%%%%%%%%%%%%%%%%%%%%
%TABLE II
%%%%%%%%%%%%%%%%%%%%%%%%%%%%%%%%%%%%%%%%
\linespread{1.0} \renewcommand{\arraystretch}{1.5}
\begin{table}[ht]
\centering
\begin{tabular}{| c | c | c | c | c |} 
\toprule[1.5pt]
$\lam_i$        & $\pxyl$        & $\pxypl$ & $\pxpyl$ & $\pxpypl$ \\
\toprule[1.5pt]
${\lam}_1$   & $\CM$ & $\BM$ & $\AM$ & $\DP$ \\
${\lam}_2$   & $\BM$ & $\CM$ & $\DP$ & $\AM$ \\
${\lam}_3$   & $\AM$ & $\DP$ & $\CM$ & $\BM$ \\
${\lam}_4$   & $\DP$ & $\AM$ & $\BM$ & $\CM$ \\
\bottomrule[1.25pt]
\end{tabular}\par
\caption{
Conditional probabilities $p (\lambda_i \vert u,v)$ for the value of the hidden variable $\lambda$ to be $\lambda_i$, for $\Mx \geq \My$. Normalization may by checked by summing the entries in each column.  The probabilities must be nonnegative, and we will see after Eqs.~\eqref{eq:ps} that the entries are nonnegative for the entire range of allowed values of $(\Mx,\My)$.}
\label{tab:general2a}
\end{table}
%%%%%%%%%%%%%%%%%%%%%%%%%%%%%%%%%%%%%%%%

For arbitrary $\Mx \geq \My$ 
in the range $0 \le \Mx \le
2$, $0 \le \My \le 2$, it follows that if we choose 
\begin{eqnarray}
  \pxone &=& \Mx/2 \ , \nonumber \\
  \pxtwo &=& \My/2  \ , \nonumber\\
  \pxthree &=& \begin{cases}
   0 & \hbox{if $\Mx + 2 \My \le 2$} \\
   %{1 \over 4} \left( \Mx + 2 \My - 2 \right) &
   \frac{1}{4} \left( \Mx + 2 \My - 2 \right) &
     \hbox{otherwise}
   \end{cases}
   \label{eq:ps}
\end{eqnarray}
in Table~\ref{tab:general2a}, then the constraints $\pxone \ge \pxtwo \ge \pxthree$
are satisfied, and Eq.~(\ref{Stable0}) simplifies to 
\begin{equation}
S = \begin{cases}
   2  + \Mx + 2\My & \hbox{if $\Mx + 2 \My \le 2$} \\
   4 &
     \hbox{otherwise} .
   \end{cases}
%S = 2 + 2 \pxone + 4 \pxtwo \, ,
%S = 2 + 2 \pxone + 4 \pxtwo - 4 \pxthree\, ,
\label{Stable1}
\end{equation}
Furthermore, one can check that for these values of $p_1$, $p_2$, and $p_3$, all the conditional probabilities in Table~\ref{tab:general2a} are nonnegative. 
It follows that, with this choice, the local deterministic model corresponding to Table~\ref{tab:general2a}  saturates the  relaxed Bell inequality in Eq.~(\ref{GeneralS}) for all values $\Mx \geq \My$. 

Finally, by symmetry, one may construct equivalent tables for the case $\My \geq \Mx$, by switching settings labels $x \leftrightarrow y$, $x' \leftrightarrow y'$ and subscripts $1 \leftrightarrow 2$ in Table~\ref{tab:general2a}. We have therefore demonstrated that the two-parameter upper bound derived in Eq.~(\ref{bound1}) is a tight upper bound, in the sense that no better bound depending only on $\Mx$ and $\My$ is possible.

It is worth noting that while the model in Tables~\ref{tab:general1a} and~\ref{tab:general2a} saturates the Hall and Banik \etal relaxed Bell inequalities in Eqs.~(\ref{HallS}) and~(\ref{BanikS}), for the respective special cases $M_1=M_2$ and $M_2=0$, the model of this section is very different from those in 
Table~I of Ref.~\cite{hall10} and Table 1 of Ref.~\cite{banik12}.
An alternative saturating model for arbitrary $M_1$ and $M_2$, that interpolates between the Hall and the Banik \etal models, is given in Appendix~\ref{sec:interpol}. However, as  will be seen below, the model in Tables~\ref{tab:general1a} and~\ref{tab:general2a} has significantly better mutual information properties.

%-----------------------------------------------------------------------------------------------------------------------------------------------
 \section{Mutual information properties and comparisons}
 \label{sec:info}
 %-----------------------------------------------------------------------------------------------------------------------------------------------

The degrees of measurement dependence, $\Mx$, $\My$, and $M$, quantify the correlation of the hidden variables $\lambda$ with Alice's settings and/or Bob's settings.
One may alternatively quantify the correlation in terms of the corresponding mutual information \cite{barrett11,hall11}, which has a more direct interpretation as the average information that may be obtained about one variable from knowledge of the other. In this section we calculate the mutual information required to achieve a violation $V$ of the CHSH inequality for the model in Tables~\ref{tab:general1a} and~\ref{tab:general2a}. We conjecture that this model is in fact optimal in the sense of requiring the lowest possible mutual information for a given violation $V$. 
 
 \subsection{Calculating mutual information}
 
The mutual information between the hidden variable $\lambda$ and the joint measurement setting $(u,v)$, measured in units of bits, is given by
\begin{eqnarray}
I & \equiv & \sum_{\lambda,u,v}p(\lambda,u,v)\log_2 \frac{p(\lambda,u,v)}{p(\lambda)p(u,v)} \nonumber \\ 
  & = & \sum_{\lambda,u,v}p(\lambda|u,v)p(u,v)\log_2 \frac{p(\lambda|u,v)}{p(\lambda)},
\label{eq:mutualinf}
\end{eqnarray}
where $p(u,v)$ is the probability distribution of joint measurement settings and $p(\lambda,u,v)\equiv p(\lambda|u,v)p(u,v)$. Note that the mutual information vanishes identically in the case of %full 
measurement independence, for which $p(\lambda|u,v)=p(\lambda)$ as per Eq.~(\ref{settingindep2}).

We will calculate the mutual information, using Eq.~(\ref{eq:mutualinf}), for the standard CHSH scenario in which the settings are chosen randomly and independently of each other (though not independently of the hidden variable). For this scenario, $p(u,v)=1/4$ for each settings pair, and the mutual information for the saturating model of the previous section simplifies to
\begin{eqnarray}
I_G = H_\Lambda -\frac{1}{4}  \left( H_{xy}+H_{xy'}+H_{x'y}+H_{x'y'}\right)\, , 
\label{IGa}
\end{eqnarray}
where 
\begin{eqnarray}
H_\Lambda \equiv -\sum_{i=1}^{4} p(\lam_i)\log_2 p(\lam_i)
\label{HpLambda}
\end{eqnarray}
and 
\begin{eqnarray}
H_{uv}\equiv -\sum_{i=1}^{4} p(\lam_i|u,v)\log_2 p(\lam_i|u,v)
\label{HpLambdaUV}
\end{eqnarray}
denote the entropies of the distributions $p(\lambda)$ and $p(\lambda|u,v)$, respectively. Note that for $p(u,v) = 1/4$, we have $p (\lambda_i) = \sum_{u,v}p(u,v)p(\lambda_i|u,v) =  \frac{1}{4} \sum_{u,v} p (\lambda_i \vert u,v)$.  

As per Table~\ref{tab:general2a}, we consider the case $\Mx \geq \My$, and restrict attention to the range $\Mx + 2 \My \leq 2$ with $\pxthree=0$. Then the CHSH inequality becomes $S \leq 2 + V_G (\Mx, \My) = 2 + \Mx + 2 \My$. This covers the whole range of possible violations $V\in [0,2]$. (As before, the case $\My \geq \Mx$ follows upon switching observer labels.) 

\begin{figure}
\centering
\begin{tabular}{@{}c@{}}
\includegraphics[width=3.25in]{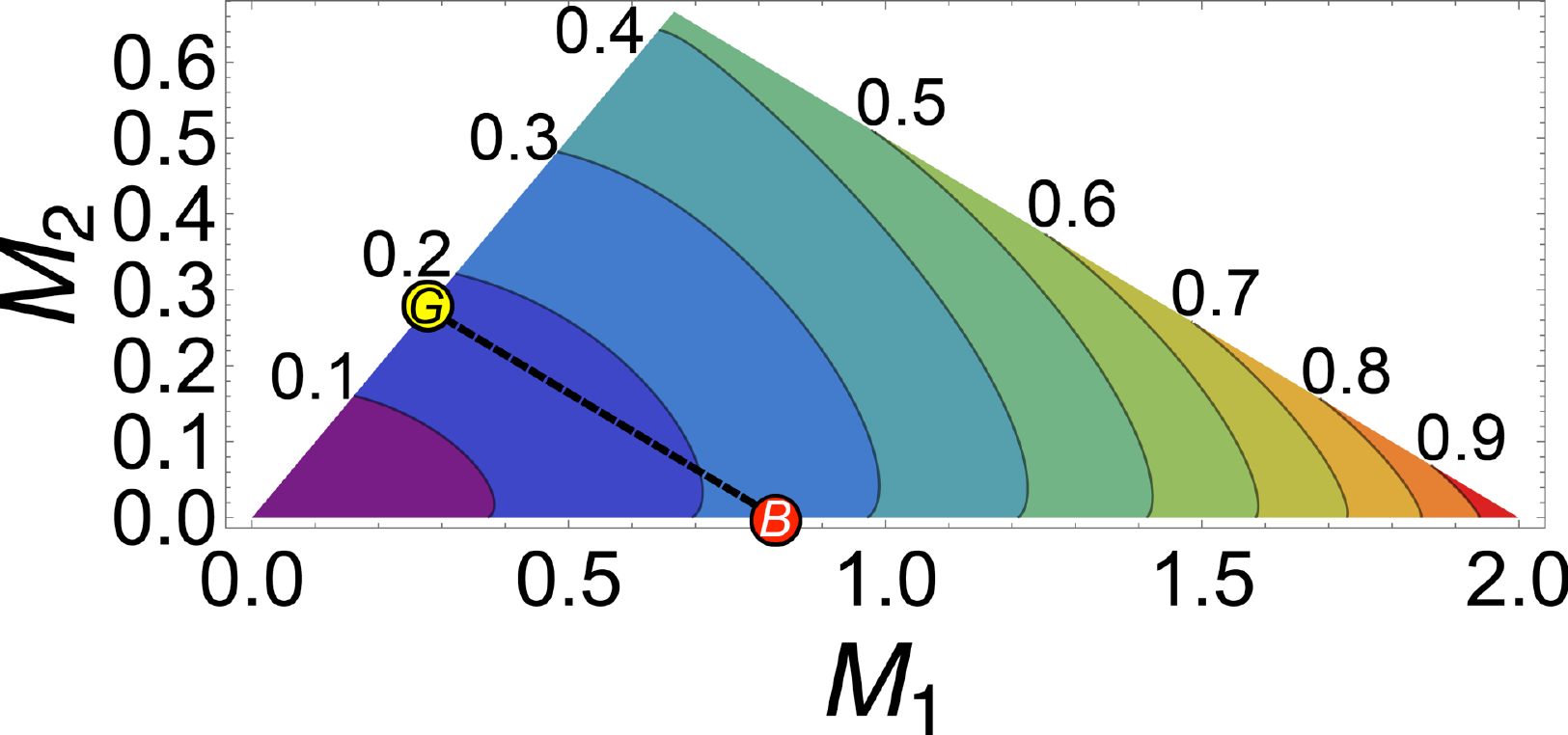} 
\end{tabular}
\caption{
\baselineskip 9pt 
``Freedom Square" plot for the region $V_G(\Mx,\My) = \Mx + 2\My \leq 2$, $\Mx \geq \My$, 
labeled by contours of mutual information $I_G(\Mx,\My) \in [0,1]$ in bits from Eq.~(\ref{eq:IGM1M2a}). 
The dashed line is the slice corresponding to the Tsirelson bound, $V_G(\Mx,\My)=\Vs$, and connects the case $(\Mx,\My)=(\Vs,0)$ (dark red circle for the Banik \etal model ($B$)) to the case $(\Mx,\My)=(\Vs/3,\Vs/3)$ (light yellow circle ($G$)). Relevant cases
are explored further in Fig.~\ref{fig:freedom_square3d}.  
The line $\Mx=\My$ 
minimizes the mutual information for the general saturating model in Tables~\ref{tab:general1a} and~\ref{tab:general2a}, for each value of Bell violation $V \in [0,2]$ (see also 
Eqs.~(\ref{eq:IGV})--(\ref{minimum}) and Fig.~\ref{fig:freedom_square3d}).
}
\label{fig:freedom_square3b}
\end{figure}

It follows, using Table~\ref{tab:general2a} and Eq.~(\ref{IGa}), and taking $\pxone=\Mx/2$, $\pxtwo=\My/2$, $\pxthree=0$ as per Eq.~(\ref{eq:ps}), that the mutual information is given by
\begin{eqnarray}
  I_{G}(\Mx,\My) & = & \frac{1}{4} \Bigg\{ 2 h\left( 1 + \frac{\Mx}{2}\right) + h\left(1-\frac{\Mx}{2} + \My\right) \nonumber \\
     & + & h\left(1-\frac{\Mx}{2} - \My\right) \Bigg\} \ ,
\label{eq:IGM1M2a}
\end{eqnarray}
where
\begin{equation}
  h(x) \equiv x \log_2(x) \ .
\end{equation}
Eq.~(\ref{eq:IGM1M2a}) for $I_G(\Mx, \My)$ is depicted in Fig.~\ref{fig:freedom_square3b}. 
%%%%%%%%%%%%%%%%%%%%%%%%%%%%%%%%%%%%%%%%
%%%%%%%%%%%%%%%%%%%%%%%%%%%%%%%%%%%%%%%%

\subsection{Comparison with previous models}

To gain further insight, and to make comparisons with previous work, it is of interest to consider the behavior of $I_G(\Mx, \My)$ for a given degree of violation $V=\Mx+2\My$ (e.g., the maximum quantum violation $\Vs$ corresponding to the dashed line in  Fig.~\ref{fig:freedom_square3b}).  For example, is the amount of mutual information minimized by choosing $\Mx=\My$ (the yellow circle in Fig.~\ref{fig:freedom_square3b} for $V=\Vs$), or by choosing $\Mx=V, \My=0$ (the red circle in Fig.~\ref{fig:freedom_square3b} for $V=\Vs$)?   And how does this minimum mutual information compare with the corresponding values for the Hall and the Banik \etal models in Refs.~\cite{hall10} and~\cite{banik12}?

First, using the relation $\Mx = V - 2\My \geq \My$,  we express Eq.~(\ref{eq:IGM1M2a}) in terms of $\My$ and the amount of violation $V$:
\begin{eqnarray}
\tilde I_{G}(V,\My)  &\equiv& I_G(V-2\My,\My) \nonumber\\ 
&=& \frac{1}{4}\left\{ 2h\Big(1 + \frac{V}{2} - \My\Big) + h\Big(1 - \frac{V}{2} + 2\My\Big) \right. \nonumber\\
& &\qquad \qquad \left.  + h\Big(1 - \frac{V}{2}\Big) \right\} \, , 
\label{IGM1M2Va}
\end{eqnarray} 
with $\My$ restricted to the range 
$0 \leq \My \leq V/3 \leq 2/3$. It is then straightforward to minimize this quantity with respect to $\My$, for any given value of the violation $V$, leading to the result 
\begin{eqnarray}
 \tilde{I}_G(V) &\equiv& \min_{\My} \tilde{I}_{G}(V,\My) =I_G(V/3,V/3)  \nonumber \\
  & = &\frac{1}{4}\Bigg\{ 3h\Big(1 + \frac{V}{6} \Big) + h\Big(1 - \frac{V}{2}\Big) \Bigg\}\, , 
\label{eq:IGV}
\end{eqnarray}
where the minimum is achieved for the values 
\begin{equation}
  \Mx = \My = M =\frac{V}{3}  \ .
  \label{minimum}
\end{equation}
 
\begin{figure}
\centering
\begin{tabular}{@{}c@{}}
\includegraphics[width=3.2in]{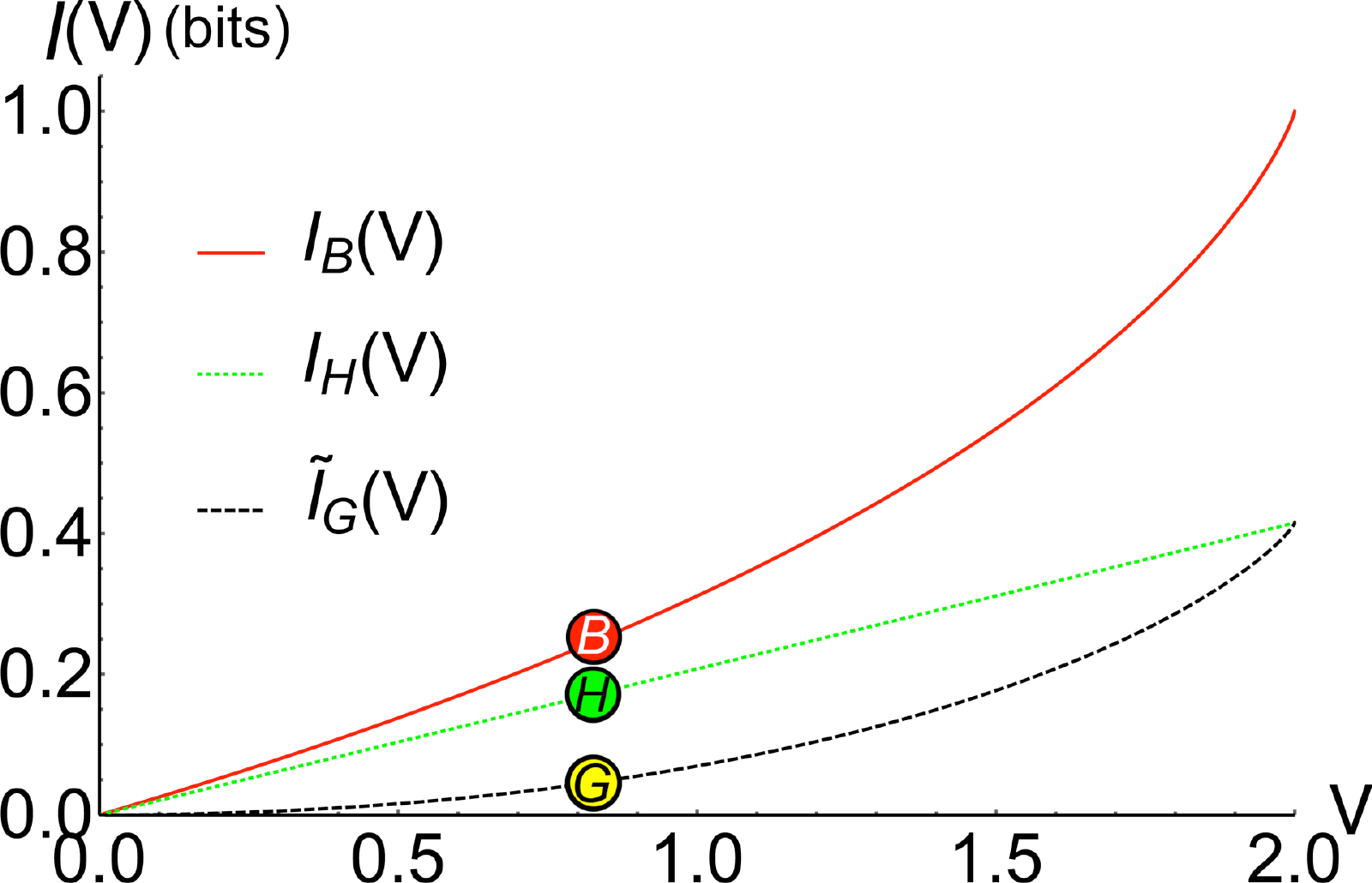} \\
\end{tabular}
\caption{
\baselineskip 9pt 
Mutual information for the Banik {\it et al.} model 
($I_B(V)$, solid red curve), the Hall model ($I_H(V)$, dotted green curve) from the literature \cite{hall10,banik12},
and the minimum of the general model 
($\tilde{I}_G(V)$, dashed black curve)
in bits, plotted as a function of CHSH violation $V\in[0,2]$.
As in Fig.~\ref{fig:freedom_square3b}, large filled
circles denote $V_G(\Mx,\My)=\Vs$.
Note that since $\tilde{I}_G(V)<I_H(V)<I_B(V)$ for all violations $0 < V < 2$,
the Hall model always requires less mutual information than the Banik model to produce a given Bell violation, while the minimum of the general model 
requires much less mutual information than the Hall or Banik models.}
\label{fig:freedom_square3d}
\end{figure}

By comparison, the Hall model from Table I of Ref.~\cite{hall10} (see also Appendix~\ref{sec:interpol}) 
has a corresponding mutual information
\begin{eqnarray}
I_H(V)
= \frac{V}{2}\log_2 \frac{4}{3},
\label{eq:Hallmutual1}
\end{eqnarray}
for a given degree of violation $V$, while the Banik {\it et al.} model
from Table 1 of Ref.~\cite{banik12} (see also Appendix~\ref{sec:interpol}) 
has a corresponding mutual information
\begin{eqnarray}
I_B(V)
    = \frac{1}{4}\Bigg\{6 + 2 h\Big(2-V\Big) - h\Big(4-V\Big) \Bigg\}\, .
\label{eq:Banikmutual1}
\end{eqnarray}
Equations~(\ref{eq:IGV}), (\ref{eq:Hallmutual1}) and~(\ref{eq:Banikmutual1}) are plotted as functions of $V$ in 
Fig.~\ref{fig:freedom_square3d},
showing that
\begin{equation}
\tilde{I}_G(V) < I_H(V) < I_B(V)
\label{eq:IGVineq}
\end{equation}
for all $V\in (0,2)$. 

%%%%%%%%%%%%%%%%%%%%%%%%%%%%%%%%%%%%%%%%
%%%%%%%%%%%%%%%%%%%%%%%%%%%%%%%%%%%%%%%%

As an example, consider the case of the maximum quantum violation $V$=$\Vs$, depicted by colored circles in Fig.~\ref{fig:freedom_square3d}.
While the Hall model requires a mutual information $I_H(\Vs) \approx 0.172$ bits (green circle ($H$)), which is less than the $I_B(\Vs)\approx 0.247$ bits required by the Banik model (red circle ($B$)), the general model of Table~\ref{tab:general2a} can achieve the maximum quantum violation with a substantially smaller mutual information,
\begin{eqnarray}
 \tilde I_G(\Vs) &=& \frac{1}{4}  \Bigg\{ 3 h\Bigg(\frac{2+\sqrt{2}}{3}\Bigg) + h(2-\sqrt{2})\Bigg\} \nonumber  \\
 & \approx & 0.0462738 \ {\rm bits}\, ,
\label{eq:IGmin}
\end{eqnarray}
(yellow circle ($G$)), corresponding via Eqs.~(\ref{Vsdef}) and~(\ref{minimum}) to 
\begin{equation}
\Mx = \My = \frac{\Vs}{3} = \frac{2}{3} \left( \sqrt{2}
     - 1 \right) \ .
     \label{mvalues}
\end{equation}
Hence, the model in Table~\ref{tab:general2a} requires significantly less mutual information between the settings and hidden variables to simulate violations of the Bell-CHSH inequality than previously studied locally causal models.

 In Appendix~\ref{sec:interpol} we construct an alternative model which also saturates the the general two-parameter bound in Eqs.~(\ref{GeneralS})--(\ref{bound1}) for arbitrary values of $M_1$ and $M_2$,  which is a simple mixture of the Hall and Banik \etal models. As is shown in Appendix~\ref{sec:info1}, this mixed model is interesting in that it requires less mutual information than either the Hall or Banik models for a given violation $V$. However, it nevertheless requires significantly more mutual information than the model in Table~\ref{tab:general2a}. 

%%%%%%%%%%%%%%%%%%%%%%%%%%%%%%%%%%%%%%%%%%%%%
\subsection{An optimality conjecture}%%%%%%%%%

Remarkably, the value of $\tilde I_G(\Vs)  \approx 0.0462738 \sim 1/22$ of a bit in Eq.~\eqref{eq:IGmin} is {\it identical} to the mutual information reported in Eq.~(37) of Ref.~\cite{hall11}, where the latter is for the local deterministic model of general singlet state correlations given in Ref.~\cite{hall10} when restricted to the CHSH scenario with detector angles chosen to maximize the quantum prediction for the Bell inequality violation.  The underlying reason for this agreement is that the hidden variable $\lambda$ for the singlet state model is represented by a point on the unit sphere, with its relation to the CHSH measurement settings wholly determined by which one of four regions of the sphere that it lies in.  In particular, these regions generate four sets of conditional probabilities that correspond to the rows of Table~\ref{tab:general2a} for $\Mx$ and $\My$ in Eq.~(\ref{mvalues}).

The full singlet state model in Ref.~\cite{hall10} has a high degree of symmetry, and requires the lowest known mutual information by far of any such model when arbitrary numbers of measurement settings are allowed on each side~\cite{hall10,hall16}. Hence, given that the saturating model in Table~\ref{tab:general2a} is similarly highly symmetric for all $M_1=M_2$ (and $p_3=0$), and requires significantly lower mutual information than other known models for general values of $V$, we conjecture that $\tilde I_G(V)$ is in fact the minimum amount of mutual information required for any locally causal model of a given Bell-CHSH violation $V$.

%-----------------------------------------------------------------------------------------------------------------------------------------------
 \section{Generalizing From Two Parameters to Four Parameters}
 \label{sec:4pbound}
 %-----------------------------------------------------------------------------------------------------------------------------------------------

So far we have been describing the degree of measurement
dependence of Alice and Bob by the traditional parameters
\cite{hall10} $\Mx$ and $\My$, as defined in general by
Eqs.~\eqref{hallMa1} and \eqref{hallMa2}, and specifically for
the CHSH scenario by Eqs.~\eqref{hallMaCHSH1} and
\eqref{hallMaCHSH2}.  There are, however, other interesting
variables that can be defined by
\begin{equation}
\Mtx  \equiv  \inf_{v\in V} \Bigg\{ \sup_{u_1,u_2\in U} \Bigg[ \int \dlam \big| p(\lam|u_1,v) - p(\lam|u_2, v) \big| \Bigg] \Bigg\}\label{Mtx} \, , \\ 
\end{equation}
\begin{equation}
\Mty \equiv \inf_{u\in U} \Bigg\{ \sup_{v_1,v_2\in V} \Bigg[ \int \dlam \big| p(\lam|u,v_1) - p(\lam|u_,v_2) \big| \Bigg]\Bigg\} \label{Mty} \, .
\end{equation}
These quantities also have a relevant physical interpretation. 
$\Mx$ describes the most serious loss of freedom that Alice (who
sets detector 1) might experience, but the actual loss of freedom
that she will experience depends on the value of $v$, the setting
on the other side.  For the general case of Eq.~\eqref{hallMa1},
the probability that she experiences this worst-case scenario
might be extremely small, if the setting $v$ that maximizes the
expression is extremely improbable.  For example, a model might
have the property that Alice can make only one choice if the
angle on the far side is between 22 degrees and $22 + 10^{-500}$
degrees, but otherwise she is completely unrestricted.  In this
case $\Mx$ would be equal to its maximal value of 2, even though
the restrictions on Alice's choices are so rare that they could
never be detected in the lifetime of many universes.  

The quantity $\Mtx$, by contrast, describes the inevitable, minimum loss
of freedom that Alice will experience, 
no matter what value the
setting $v$ has.  For the example just discussed, $\Mtx$ would
equal zero.  $\Mtx$ could of course also be misleading, since
again the setting $v$ that minimizes the expression in
Eq.~\eqref{Mtx} might be extremely improbable.  Nonetheless, we
can always count on $\Mtx$ and $\Mx$ to bracket the degree of
measurement dependence that Alice will experience.  In the
context of our \two-setting CHSH Bell test, the difference between
$\Mtx$ and $\Mty$ and the usual $\Mx$ and $\My$ is the choice
between taking the $\min$ or the $\max$ of the two quantities on
the right-hand side of Eqs.~\eqref{hallMaCHSH1} and
\eqref{hallMaCHSH2}.

Finally, we may also interpret $\Mtx$ in terms of an experiment in which 
one tries to use a measurement of $\lambda$ to distinguish between two of Alice's measurement settings, with Bob's setting the same in both cases.  Varying over Bob's setting and Alice's two settings, the minimum value of the probability that 
one will be able to identify Alice's setting is given by $\half \left( 1 + \frac{1}{2} \Mtx\right)$ (see Sec.~III). A similar interpretation applies to $\half \left( 1 + \frac{1}{2} \Mty\right)$.

\subsection{The general four-parameter bound}

If we reexamine the proof of the general \two-parameter bound in
Sec.~\ref{sec:general}, we 
find that it leads not only to the
\two-parameter bound of Eq.~\eqref{bellT1s9}, but also to a more
detailed \four-parameter bound, based on $\Mx$, $\My$, $\Mtx$, and
$\Mty$.  To see this, start by noticing that the quantities $T_2$
and $T_3$, defined in Eqs.~\eqref{bellT2s} and \eqref{bellT3s},
are identical to the two quantities appearing inside the curly
brackets in the expression for 
$\My$ in Eq.~\eqref{hallMaCHSH2}. 
The larger of these two quantities becomes $\My$, but the smaller
becomes $\Mty$, as can be seen from the definition in
Eq.~\eqref{Mty}.  Thus,
\begin{equation}
T_2 + T_3 = \My + \Mty \, .
\end{equation}

From Eqs.~\eqref{bellT1s3}--\eqref{bellT1s5a}, we can conclude
that
\begin{equation}
T_1 \le 2 + \Mx[y] \, ,
\label{T1-4p1}
\end{equation}
where we introduce the definitions
\begin{eqnarray}
\Mx[v] &\equiv&  \int d\lam \Big| p(\lam | \x, v)  - p(\lam | 
   \xp , v)  \Big| \, , \label{Mx[v]} \\
\My[u] &\equiv&  \int d\lam \Big| p(\lam | u, \y)  - p(\lam | 
   u , \yp)  \Big| \, , \label{My[u]}
\end{eqnarray}
where $u \in \{\x,\xp\}$ and $v \in \{\y,\yp\}$.  But there is
nothing about this system that makes any absolute distinction
between $\y$ and $\yp$, so one could have constructed a
rearrangement of the derivation shown, in which $\yp$ would
appear in Eqs.~\eqref{bellT1s3}--\eqref{bellT1s5a}, instead of
$\y$.  Then, in addition to Eq.~\eqref{T1-4p1}, we would also have
\begin{equation}
T_1 \le 2 + \Mx[y'] \, .
\label{T1-4p2}
\end{equation}
From Eqs.~\eqref{T1-4p1} and \eqref{T1-4p2}, we conclude that
\begin{equation}
T_1 \le 2 + \Mtx \, ,
\end{equation}
and then finally
\begin{equation}
\ES \le 2 + \My + \Mty + \Mtx \, .
\label{bell4p-T1s6}
\end{equation}

The claim that we can interchange $\y$ and $\yp$ is not
completely obvious, because the definition of $S$, in
Eq.~\eqref{bellE}, is not invariant under $\y \leftrightarrow
\yp$.  The change in $S$, however, can be compensated by
redefinitions of the outcome variables $\yb$ and $\ybp$, so the
result shown in Eq.~\eqref{T1-4p2} is correct. Probably the
easiest way to show this clearly is to explicitly construct the
rearrangement of the original derivation, which we do in Appendix~\ref{sec:4boundproof}, to derive Eq.~\eqref{T1-4p2}.

Now, following the original derivation in Sec.~\ref{sec:general},
we use the fact that the formalism makes no distinction between
the first and second observer's detectors, so we can carry out a
parallel derivation reversing the treatment of $\x$ and $\y$, and
hence 1 and 2, showing that
\begin{equation}
\ES \le 2 + \Mx + \Mtx + \Mty \, .
\label{bell4p-T1s7}
\end{equation}
Since Eqs.~\eqref{bell4p-T1s6} and \eqref{bell4p-T1s7} are both
valid inequalties, $\ES$ must be bounded by the smaller of the
two, and of course it must be bounded by 4.  So, finally,
\begin{equation}
\ES \le 2 + \min \Big\{\Mtx + \Mty + \min\{\Mx,\My\},\, 2  \Big\} \, .
\label{bell4p-T1s9}
\end{equation}
We will refer to this equation as the general \four-parameter bound. 

Note that the general \four-parameter bound immediately implies
the general \two-parameter bound of Eq.~\eqref{bellT1s9}, since
$\Mtx \le \Mx$, and $\Mty \le \My$.  But, for any model where
$\Mtx \not = \Mx$, or $\Mty \not = \My$, the \four-parameter bound
will be tighter than the \two-parameter bound.  This statement, of
course, does not contradict our previous statement that the
\two-parameter bound is tight --- it is tight, in the sense that
it is not possible to have a more stringent bound that depends
only on the parameters $\Mx$ and $\My$.  But with the additional
information involved in specifying $\Mtx$ and $\Mty$, the more
stringent bound of Eq.~\eqref{bell4p-T1s9} can be established.

\subsection{Saturating the \four-parameter bound}

Given that the general \four-parameter bound is more stringent
than the \two-parameter bound, we should ask whether the
\four-parameter bound is tight.  We follow the same procedure that
we used in Sec.~\ref{sec:model}, showing in this case that for
each allowed value of $(\Mx, \My, \Mtx, \Mty)$, at least one
consistent model exists that saturates the bound.

In this case the construction of the model is more complicated. 
The model described in Tables~\ref{tab:general1a} and
\ref{tab:general2a} was found essentially by trial and error, but
it is much harder when there are four independent parameters. 
However, by examining the proof of the bound, step by step, it is
possible to list exactly what properties the conditional
probabilities must obey for the bound to be saturated.  These
properties do not determine the conditional probabilities
uniquely, but they constrain the system enough so that we were
then able to use trial and error methods to construct a general
\four-parameter model, for any allowed $(\Mx, \My, \Mtx, \Mty)$,
which saturates the bound and thereby proves that the bound is
tight: it is not possible to have a more stringent bound that
depends only on the parameters $\Mx$, $\My$, $\Mtx$, and $\Mty$. 
The \four-parameter model that we will present has the property
that it reduces to the \two-parameter model of
Tables~\ref{tab:general1a} and \ref{tab:general2a} when
$\Mtx \rightarrow \Mx$ and $\Mty \rightarrow
\My$. Here we describe the \four-parameter model, and in
Appendix~\ref{sec:4pmodel} we will summarize the details of the
construction.

The allowed range of variables is of course restricted by 
\begin{equation}
\Mx, \My, \Mtx, \Mty \in [0,2] \ , \ \Mtx \le \Mx \ , \ \Mty \le
\My \, ,
\label{const1}
\end{equation}
but with four parameters there is also a triangle inequality that
limits the amount by which $\Mx$ and $\Mtx$ can differ, and
similarly for $\My$ and $\Mty$.  Specifically,
\vbox{}\pagebreak
\begin{eqnarray}
\Mx &=& \sum_{i=1}^4 \, \Bigl| \pli{x}{y} - \pli{x'}{y} \Bigr| \nonumber\\
&=& \sum_{i=1}^4 \, \Bigl| \bigl[\pli{x}{y} - \pli{x}{y'}\bigr] \nonumber \\
&& \qquad + \bigl[\pli{x}{y'} - \pli{x'}{y'}\bigr] \nonumber\\
&& \qquad + \bigl[\pli{x'}{y'}) - \pli{x'}{y}\bigr] \Bigr| \nonumber\\
&\le& \sum_{i=1}^4 \, \Bigl| \bigl[\pli{x}{y} - \pli{x}{y'}\bigr] \Bigl| 
   \nonumber \\
&& \qquad + \Bigl| \bigl[\pli{x}{y'} - \pli{x'}{y'}\bigr] \Bigr| \nonumber\\
&& \qquad + \Bigl| \bigl[\pli{x'}{y'}) - \pli{x'}{y}\bigr] \Bigr| \nonumber\\
&=& \My + \Mtx + \Mty \ .
\end{eqnarray}
There is a parallel identity that can be derived by interchanging
1 and 2, so we have
\begin{equation}
  \Mx - \Mtx \le \My + \Mty \ , \quad \My - \Mty \le \Mx +
     \Mtx \ .
\label{const2}
\end{equation}
Equations \eqref{const1} and \eqref{const2} define the allowed
range of variables, except that we will also, without loss of
generality, adopt the convention that $\Mx \ge \My$.  (If this is
not the case, the labels 1 and 2 can be interchanged.)

Table~\ref{tab:general1a} can be used again, but we need a new
table of conditional probabilities to replace
Table~\ref{tab:general2a}.  In principle one table of conditional
probabilities would suffice, but the individual entries become
rather complicated, so we instead first introduce
Table~\ref{tab:4param1}, which describes the model only for the
restricted case of $\My + \Mtx + \Mty \le 2$.

\def\qone{q_1}
\def\qtwo{q_2}
\def\qthree{q_3}
\def\qfour{q_4}
\def\Rf{R}
\def\Rfbar{\bar R}

\begin{widetext}

%%%%%%%%%%%%%%%%%%%%%%%%%%%%%%%%%%%%%%%%
%TABLE III: 4 parameters, \bar M_1 + \bar M_2 + M_2 \le 2
%%%%%%%%%%%%%%%%%%%%%%%%%%%%%%%%%%%%%%%%
\linespread{1.0} \renewcommand{\arraystretch}{1.5}
\begin{center}
\begin{table}[ht]
\centering
\begin{tabular}{| c | c | c | c | c |} 
\toprule[1.5pt]
$\lam_i$        & $\pxyl$        & $\pxypl$ & $\pxpyl$ & $\pxpypl$ \\
\toprule[1.5pt]
${\lam}_1$ &
$\qone{\text +}\frac{1}{4}({\My{\text +}\Mtx{\text +}\qtwo})$ &
$\qone{\text +}\frac{1}{4}({\My{\text +}\Mtx{\text -}\qtwo})$ &
$\qone{\text +}\frac{1}{2}({{\text -}\Mx{\text +}\Mtx{\text +}\Mty{\text +}\qtwo})$ &
$\qone$ \\
${\lam}_2$ &
$\qone{\text +}\frac{1}{4}({{\text -}\My{\text +}\Mtx{\text +}2 \Mty{\text +}\qtwo})$ &
$\qone{\text +}\frac{1}{4}({{\text -}\My{\text +}\Mtx{\text +}2 \Mty{\text +}\qtwo})$ &
$\qone$ &
$\qone{\text +}\frac{1}{2}{\Mty}$ \\
${\lam}_3$ &
$\qone{\text +}\frac{1}{2}({\My{\text -}\qtwo})$ &
$\qone$ &
$\qone{\text +}\frac{1}{4}({2 \Mx{\text +}\My{\text -}\Mtx{\text -}3 \qtwo})$ &
$\qone{\text +}\frac{1}{4}({\My{\text +}\Mtx{\text -}\qtwo})$ \\
${\lam}_4$ &
$\qone$ &
$\qone{\text +}\frac{1}{2}{\My}$ &
$\qone{\text +}\frac{1}{4}({\My{\text +}\Mtx{\text +}\qtwo})$ &
$\qone{\text +}\frac{1}{4}({\My{\text +}\Mtx{\text +}\qtwo})$ \\
\bottomrule[1.25pt]
\end{tabular}\par
\caption{
Conditional probabilities $p (\lambda_i \vert u,v)$ for the value of the hidden variable $\lambda$ to be $\lambda_i$, for $\Mx \ge \My$ and
\hbox{$\My + \Mtx + \Mty \le 2$.}
\label{tab:4param1}
}
\end{table}
\end{center}
%%%%%%%%%%%%%%%%%%%%%%%%%%%%%%%%%%%%%%%%

\vskip -20pt

Here
\begin{eqnarray}
\qone &\equiv& \frac{1}{8} (2 - \My - \Mtx - \Mty) \, ,
\label{qone}\\
\qtwo &\equiv& \min(\Mx - \Mtx, \My) \, .
\label{qtwo}
\end{eqnarray}

When $\My + \Mtx + \Mty > 2$, there are additional terms that
need to be added, as shown in Table~\ref{tab:4param2}.

%%%%%%%%%%%%%%%%%%%%%%%%%%%%%%%%%%%%%%%%
%TABLE IV: 4 parameters, Arbitrary \bar M_1 + \bar M_2 + M_2.
%%%%%%%%%%%%%%%%%%%%%%%%%%%%%%%%%%%%%%%%
\linespread{1.0} \renewcommand{\arraystretch}{1.5}
\begin{table}[ht]
\centering
\begin{tabular}{| c | c | c | c | c |} 
\toprule[1.5pt]
$\lam_i$        & $\pxyl$        & $\pxypl$ & $\pxpyl$ & $\pxpypl$ \\
\toprule[1.5pt]
$\lambda_1$ &
$P^{(0)}_{1,1} + \qthree $ &
$P^{(0)}_{1,2} - \qthree - \qfour $ &
$P^{(0)}_{1,3} - \qthree + \qfour $ &
$P^{(0)}_{1,4} + \qthree $ \\
$\lambda_2$ &
$P^{(0)}_{2,1} -\qthree + \qfour $ &
$P^{(0)}_{2,2} + \qthree + 2 \qfour $ &
$P^{(0)}_{2,3} + \qthree $ &
$P^{(0)}_{2,4} - \qthree + \qfour $ \\
$\lambda_3$ &
$P^{(0)}_{3,1} - \qthree - \qfour $ &
$P^{(0)}_{3,2} + \qthree $ &
$P^{(0)}_{3,3} + \qthree - 2 \qfour $ &
$P^{(0)}_{3,4} - \qthree - \qfour $ \\
$\lambda_4$ &
$P^{(0)}_{4,1} + \qthree $ &
$P^{(0)}_{4,2} - \qthree - \qfour $ &
$P^{(0)}_{4,3} - \qthree + \qfour $ &
$P^{(0)}_{4,4} + \qthree $ \\
\bottomrule[1.25pt]
\end{tabular}\par
\caption{
Conditional probabilities $p (\lambda_i \vert u,v)$ for the value of the hidden variable $\lambda$ to be $\lambda_i$, for any allowed values of $\Mx$, $\My$, $\Mtx$,
and $\Mty$, provided that $\Mx \ge \My$.  Here
$P^{(0)}_{i,j}$ refers to the corresponding entries of Table
\ref{tab:4param1}. }
\label{tab:4param2}
\end{table}
%%%%%%%%%%%%%%%%%%%%%%%%%%%%%%%%%%%%%%%%

\newbox\spacebox
\setbox\spacebox = \hbox{$\frac{1}{4} \big[-2 - \Mtx - \Mty +
     \min(\Mx + \Mty,2) + \max(\My + \Mtx,2) - \qtwo \big]$}

The functions $\qthree$ and $\qfour$ vanish for $\My + \Mtx +
\Mty \le 2$, and they are given in general by
\pagebreak
\begin{align}
\qthree &= \begin{cases} 
     \hbox to \wd\spacebox{0\hfill} & \hbox{if $\My + \Mtx + \Mty
                                 \le 2 \ ,$}\\
     \frac{1}{8} \big[\My + \Mtx + \Mty - 2 \big] \, , &
     \hbox{otherwise ,} 
     \end{cases}
     \label{qthree}\\
\qfour &= \begin{cases}
     0 & \hbox{if } \My + \Mtx + \Mty \le 2 \, ,\\
     \frac{1}{4} \big[-2 - \Mtx - \Mty + \min(\Mx + \Mty,2)
     + \max(\My + \Mtx,2) - \qtwo \big] &
     \hbox{otherwise \, .} 
   \end{cases}
   \label{qfour}
\end{align}

\end{widetext}

For $\My + \Mtx + \Mty > 2$, the function $\qfour$ can also be
written as
\begin{equation}
\begin{aligned}
&\frac{1}{4} \big[ \max(\Rfbar - \Rf, \My) +
     \max(\Rf,\Mty) \\
   & \qquad - \max(\Rfbar, \My) - \Mty \big] \, ,
\end{aligned}
\end{equation}
where
\begin{equation}
\begin{aligned}
\Rf &\equiv \My + \Mtx + \Mty - 2 \, ,\\
\Rfbar &\equiv \Mx + \My + \Mty - 2 \, ,
\label{Rdefs}
\end{aligned}
\end{equation}
from which it can be easily seen that $\qfour$ has two significant
properties: (1) when $\Mtx = \Mx$ and $\Mty = \My$, $\qfour$
vanishes, which allows one to see that the entire solution
reduces to the two-parameter solution in that case; (2) $\qfour$
and $\qthree$ both vanish when $\My + \Mtx + \Mty = 2$, which
assures that these function are continuous at $\My + \Mtx + \Mty
= 2$.  (Continuity is not required, but is desirable on grounds
of simplicity.) 

To verify that this model has the required properties, one must
verify that
\begin{equation}
  \Mx[\y] = \Mx \ , \ \Mx[\yp] = \Mtx \ , \ \My[\x] = \My \ , \
     \My[\xp] = \Mty \ ,
  \label{Mvalues}
\end{equation}
where $\Mx[v]$ and $\My[u]$ were defined by Eqs.~\eqref{Mx[v]}
and \eqref{My[u]}, that
\begin{equation}
\sum_{i=1}^4 p(\lambda_i | u, v) = 1
\label{4pnorm}
\end{equation}
for all $u \in \{\x, \xp\}$, and $v \in \{\y, \yp \}$, that
\begin{equation}
0 \le p(\lambda_i | u, v) \le 1 \ ,
\end{equation}
for all $i \in \{1, 2, 3, 4\}$, $u \in \{\x, \xp\}$, and $v \in
\{\y, \yp \}$, and finally that
\begin{equation}
S = 2 + \min \Big\{\Mtx + \Mty + \min\{\Mx,\My\},\, 2  \Big\} \, .
\end{equation}
The verification of these properties, which depends on keeping in
mind the restrictions of Eqs.~\eqref{const1} and \eqref{const2}
and the convention $\Mx \ge \My$, is tedious but straightforward.

\subsection{Mutual information of the four-parameter model}

\begin{figure}
\centering
\begin{tabular}{@{}c@{}}
\includegraphics[width=3.4in]{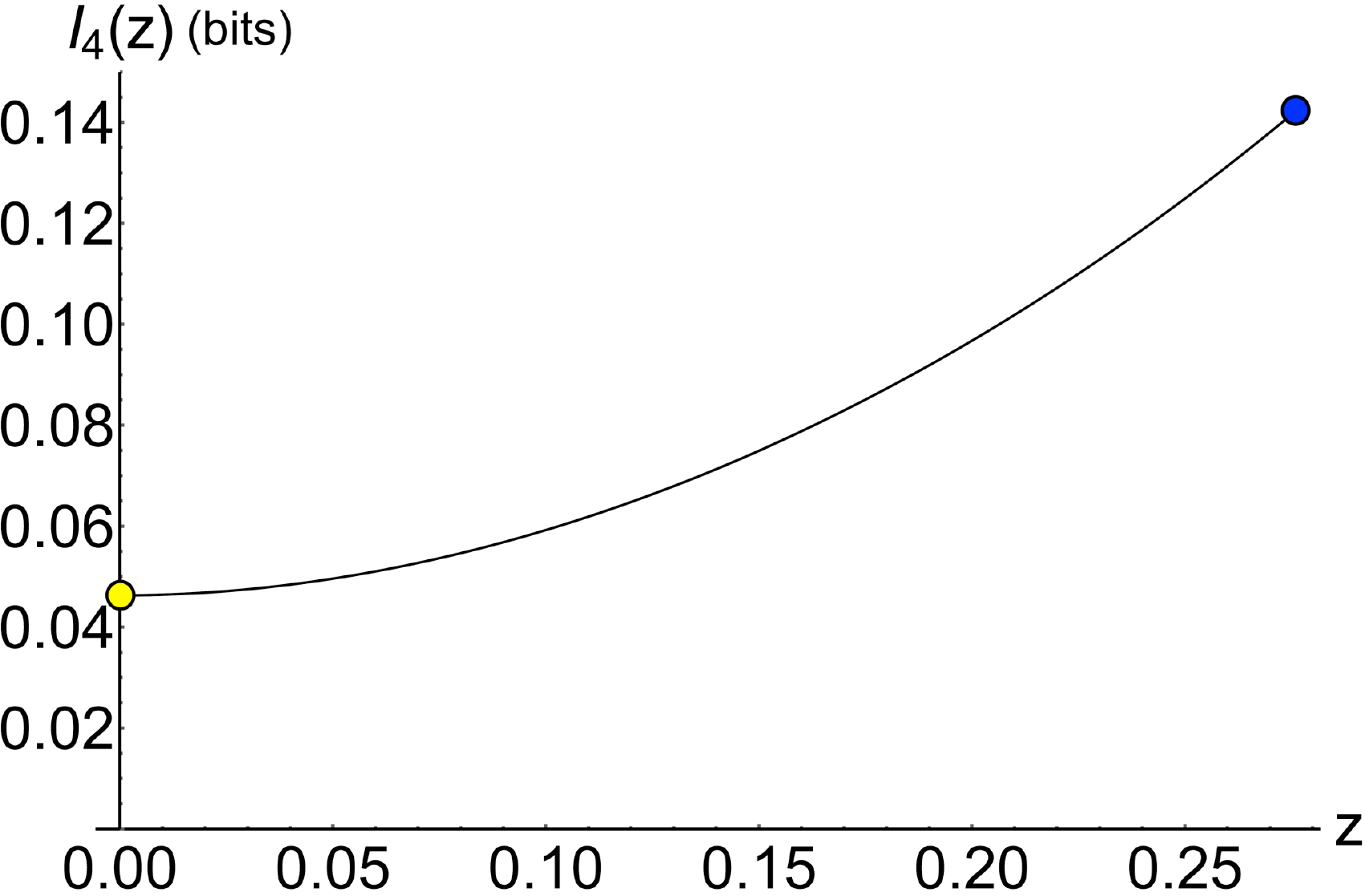} \\
\end{tabular}
\caption{Plot of mutual information $I_4(\z)$, in bits, for the four-parameter model in Table~\ref{tab:4param1}, for $\Mx \ge \My$ and
$\My + \Mtx + \Mty \le 2$. $I_4(\z)$, given by Eq.~\eqref{eq:mutinfI4},
is calculated for $\Mx = \My = \Vs/3 + 2
\z$, $\Mtx = \Mty = \Vs/3 - \z$, so in all cases $S = 2 + \Vs$. The mutual information $I_4(\z)$ is minimized via Eq.~\eqref{eq:mutinfI4min} when $I_4(0)\approx 0.0463$ bits (light yellow circle) and maximized when $I_4(\Vs/3) \approx 0.1423$ bits (dark blue circle).
}
\label{fig:mbar-ne-m}
\end{figure}

Since the \four-parameter model reduces to the \two-parameter model
when $\Mtx = \Mx$ and $\Mty = \My$,
it reproduces the \two-parameter solution for $\Mx = \My = \Vs/3$,
which gives the quantum violation of the Bell-CHSH inequality
(Tsirelson bound) with a very low mutual information, $\approx$ 0.0463
bits, as per Eq.~(\ref{eq:IGmin}). To show one example of how the mutual information changes
when $\Mtx \not = \Mx$ or $\Mty \not = \My$, we show in
Fig.~\ref{fig:mbar-ne-m} a plot of the mutual information of the
\four-parameter model as a function of $\z$, where $\Mx = \My =
\Vs/3 + 2 \z$ and $\Mtx = \Mty = \Vs/3 - \z$, so in all cases $S =
2 + \Vs$.  
In this case, the mutual information $I_4(\z)$ is given by
\begin{eqnarray}
I_4(\z) & = & 1 + \frac{3 \z}{2} +
h\Bigg(\frac{2-\sqrt{2}}{4}\Bigg) 
\nonumber \\
 & + & 2 h\Bigg(\frac{2+\sqrt{2}-6\z}{12}\Bigg) +  h\Bigg(\frac{2+\sqrt{2}+12\z}{12}\Bigg) \nonumber \\
 & - &  2 h\Bigg(\frac{2-3\z}{8}\Bigg) -\frac{1}{4}h\Big(1+3\z\Big)
\label{eq:mutinfI4}    
\end{eqnarray}
The mutual information in Eq.~\eqref{eq:mutinfI4} is minimized (light yellow circle in Fig.~\ref{fig:mbar-ne-m}) when $\z=0$, 
yielding 
\begin{equation}
I_4(\z=0) = \tilde I_G(\Vs) \approx 0.0462738 \ {\rm bits,}
\label{eq:mutinfI4min}
\end{equation}
which is identical to the value of $\tilde I_G(\Vs)$ in
Eq.~\eqref{eq:IGmin}. $I_4(\z)$
grows monotonically with $\z$, to a maximum value of $\approx$ 0.1423 bits (dark blue circle in Fig.~\ref{fig:mbar-ne-m}) when
$\z = \Vs/3$ $\approx$ 0.2761.

%%%%%%%%%%%%%%%%%%%%%%%%%%%%%%%%%%%%%%%%
%%%%%%%%%%%%%%%%%%%%%%%%%%%%%%%%%%%%%%%%

%-----------------------------------------------------------------------------------------------------------------------------------------------
 \section{Discussion}
 \label{sec:conc}
%-----------------------------------------------------------------------------------------------------------------------------------------------

As recognized by Bell himself, the measurement-independence (or freedom-of-choice) assumption is crucial to the derivation of Bell's theorem \cite{,bell76,bell87}. Relaxing this assumption leads to a potent loophole in the theorem, and opens space for families of locally causal hidden-variable models that could reproduce the quantum predictions for entangled states.
As experimental efforts to address the measurement-independence loophole in tests of Bell's inequality continue to improve \cite{scheidl10,gallicchio14,aktas15,handsteiner17,BigBellTest2,rauch18,li18}, it is therefore critical to investigate properties of locally causal models, distinct from quantum mechanics, that could exploit such a loophole to remain viable in the face of various tests.

Building on  work in Refs.~\cite{hall10,hall11,banik12}, we have constructed a general framework for relaxing the 
measurement-independence assumption for two-particle tests of Bell's inequality, to accommodate arbitrary amounts of reduced experimental freedom for each observer while satisfying local causality. This framework allows for interpolation between previously studied symmetric models, in which each observer gives up the same amount of freedom \cite{hall10,hall11}, and one-sided models, in which one observer gives up some freedom while the other maintains perfect freedom \cite{banik12}. We have derived 
two new, relaxed Bell-CHSH inequalities
for this general framework, 
which subsume previously studied models as special cases of our more general 
\two- and \four-parameter bounds. We show that these bounds are tight
by providing local deterministic models which saturate 
each bound for all regimes of measurement dependence for each observer.

We have also calculated the efficiency of these saturating models for simulating Bell-CHSH violations, as measured by mutual information between the Bell-test detector settings and any hidden variables that affect measurement outcomes. Most interestingly, we find that the \two- and \four-parameter models in our
Tables~\ref{tab:general2a} and~\ref{tab:4param1}
are very efficient, capable of achieving a given violation of the Bell-CHSH inequality with far less mutual information between the hidden variables and the joint detector settings than is needed by locally causal models that had previously been identified in the literature. We conjecture that our models are optimal in the sense that they achieve (for $\Mx = \My = \Mtx = \Mty$) the minimum possible mutual information for a given Bell-CHSH violation. Although the interpolating model in Table~\ref{tab:general2} of Appendix~\ref{sec:interpol} is not optimal compared to Tables~\ref{tab:general2a} and~\ref{tab:4param1}, it too requires less mutual information than previously studied models, which it specifically reproduces as special cases.

For each model in this class, we find that only a comparatively small degree of measurement dependence (as measured in bits of mutual information) must be assumed
in order to reproduce the predictions from quantum theory, compared to
hidden-variable models that exploit other loopholes such as the locality or communication loophole \cite{toner03} or models that relax determinism \cite{hall11}. 

Our framework for considering such models is quite general. For example, while measurement-dependent models allow correlations between the measurement settings and  $\lambda$, our framework makes no stipulations about where or when in space-time the hidden variable $\lambda$ is created or becomes relevant;  
indeed, our formalism remains agnostic about whether $\lambda$ represents 
degrees of freedom associated with specific space-time events at all. For example, $\lambda$ could, in principle, be associated with entire space-time regions or hypersurfaces \cite{wiseman14}, or with even more fundamental degrees of freedom from which classical space-time (consistent with general relativity) might emerge. 
Note that the results in this work also apply to stochastic models, and thus are consistent with---but do not require the assumption of---determinism.

We note in particular that relaxing the measurement-independence assumption 
does not require the additional assumption of ``superdeterminism," although the two have at times been conflated in the literature \cite{thooft07,hossenfelder11,hossenfelder12,hossenfelder14,thooft12,thooft14,thooft14b,thooft14c,thooft17}.  
For concreteness, we consider the definition of superdeterminism used by 't Hooft \cite{thooft14}: {\it ``Superdeterminism may be defined to imply that not only all physical phenomena are declared to be direct consequences of physical laws that do not leave anything anywhere to chance (which we refer to as `determinism'), but it also emphasizes that the observers themselves behave in accordance with the same laws. \ldots
The fact that an observer cannot reset his or her measuring device without changing physical states in the past is usually thought not to be relevant for our description of physical laws." }
He further argues, with regard to the correlations between past physical states and present measurement choices \cite{thooft14}:
{\it ``We claim that not only there are correlations, but the correlations are also extremely strong."
}

Although the phrase ``extremely strong"  
is only a qualitative statement, one might interpret this claim to mean that in a superdeterministic universe, Alice and Bob would have no freedom whatsoever to choose Bell-test measurement settings, corresponding to $M=2$ from Eq.~(\ref{hallMa}) and thus $F=1-\frac{M}{2}=0$ in Eq.~(\ref{freedom}). Such a maximally deterministic model has been presented by Brans~\cite{brans88}. In contrast to such an ``extreme'' case of superdeterminism, we note that locally causal models 
that exploit the measurement-independence
loophole, of the sort analyzed here, require quite modest amounts of reduced experimental freedom---as measured by $M$, $\Mx$, and $\My$, or by information-theoretic measures such as mutual information---in order to mimic the predictions from quantum mechanics.
(See also Refs.~\cite{hall10,hall11,hall16,banik12}.) In short, relaxation of experimenters' freedom of choice need not be an ``all or nothing" assumption. While superdeterminism represents one logical possibility for how the measurement-independence 
assumption can be relaxed, it is not the only such possibility. 

%-----------------------------------------------------------------------------------------------------------------------------------------------
 \section{Conclusions}
 \label{sec:conc1}
%-----------------------------------------------------------------------------------------------------------------------------------------------

In this work, we have derived two new, relaxed Bell-CHSH inqualities within a general framework where the assumption of measurement independence can be relaxed to independent degrees for both observers.

In future work, it would be interesting to investigate models of the singlet state compatible with our general bound, generalizing those presented in Refs.~\cite{hall10,hall11,banik12,barrett11}, to determine whether there exist locally causal models that can produce the same amount of Bell violation for smaller values of the measurement-dependence parameters $M$, and/or that would require less mutual information between joint detector settings and hidden variables.
It would be of additional interest to further explore whether the results in this work could be generalized to other Bell inequalities beyond Bell-CHSH, for example, those with more than two measurement settings per observer, or those which are not symmetric under correlated flips of the measurement outcomes.
The approach we have developed here could also be generalized to locally causal models of correlations among 
$N$-particle ``GHZ" entangled states (with $N > 2$) \cite{greenberger90,mermin90b}. 
Finally, it would be of interest to
develop a deeper understanding of this family of 
locally causal models that relax the measurement-independence assumption
in terms of causal space-time structure (e.g. \cite{friedman13a}).

%----------------------------------------------------------------------------------------------------------------------------------------------------------------------------------
{\it Acknowledgements.---} 
The authors would like to thank Anton Zeilinger, Thomas Scheidl, Johannes Handsteiner, Dominik Rauch, Calvin Leung, and David Leon, along with the two anonymous referees for helpful discussions.
A.S.F. would like to dedicate this paper to his brother, Barry J. Friedman (1981-2018). A.S.F, A.H.G., D.I.K., and J.G. acknowledge support from NSF INSPIRE Award PHY-1541160. Portions of this work were conducted in MIT's Center for Theoretical Physics and supported in part by the U.S. Department of Energy under Contract No.~DE-SC0012567. M.J.W.H. acknowledges the support of the Australian Research Council Centre of Excellence CE110001027. 
%----------------------------------------------------------------------------------------------------------------------------------------------------------------------------------

\appendix

\section{Interpolating Between CHSH Model Tables From Previous Work}
\label{sec:interpol}

In this Appendix we construct another model that saturates the general \two-parameter bound of Eqs.~(\ref{GeneralS}) and~(\ref{bound1}), while at the same time interpolating between the models in Table 1 of Banik \etal \cite{banik12} and Table I of Hall \cite{hall10} in the physically significant region of parameter space corresponding to $M_1 + M_2 + {\rm min} \{ M_1, M_2 \} \leq 2$. At its optimum parameters within this region, the interpolating model requires less mutual information between the hidden variables and detector settings than either the Hall or Banik \etal models, although it requires significantly more mutual information than the \two-parameter model of Sec.~\ref{sec:model}\null.
In the region for which $M_1 + M_2 + {\rm min} \{ M_1, M_2 \} > 2$, where the Banik \etal model does not exist, the interpolating model generalizes the Hall model, reducing to Table II of Hall \cite{hall10} when $M_1 = M_2$.

\begin{figure}
\centering
\begin{tabular}{@{}c@{}}
\includegraphics[width=2.5in]{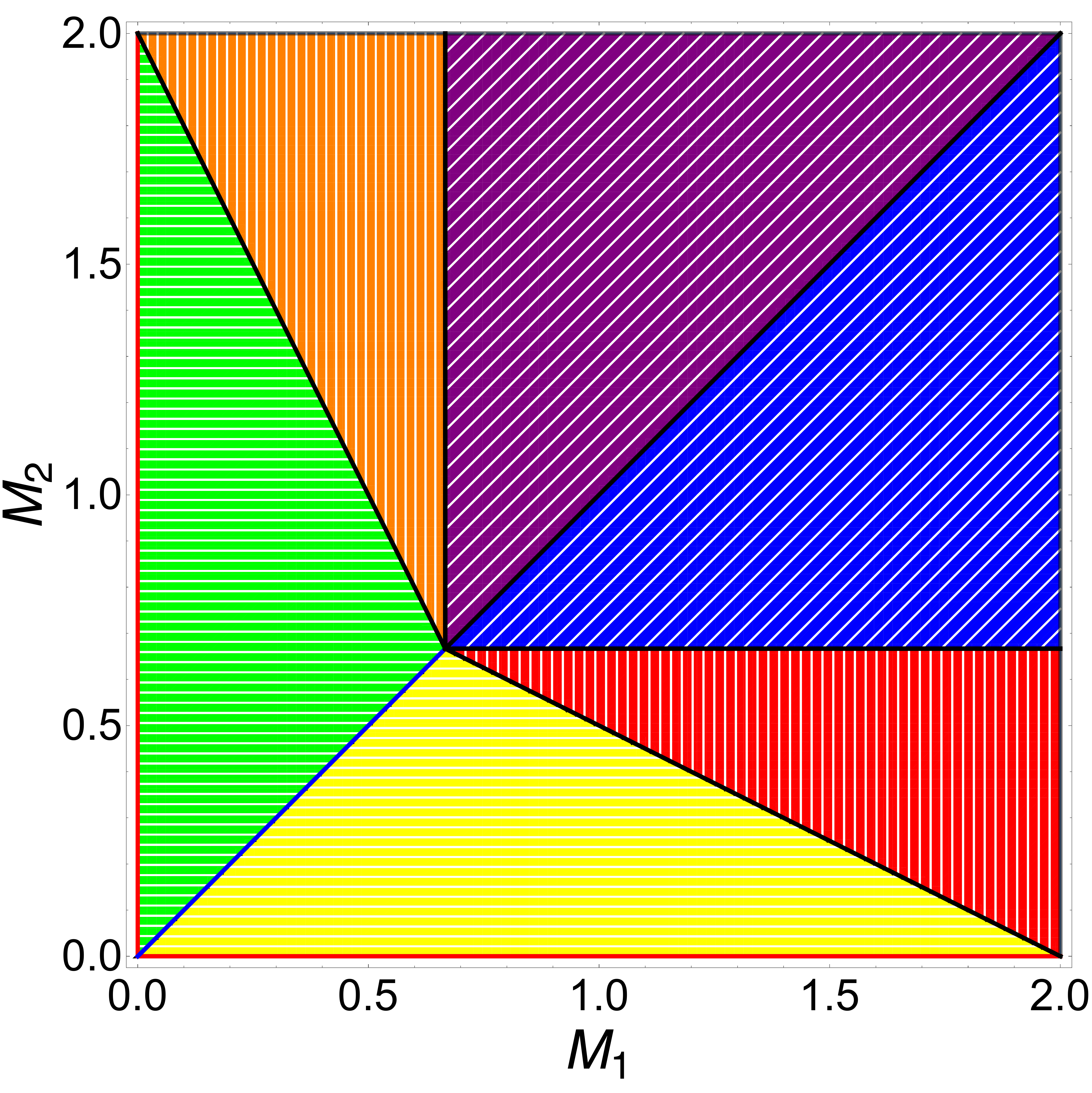} %&
\vspace{-0.5cm}
\end{tabular}
\caption{
We construct an ``interpolating'' model for $p (\lambda_i \vert
u, v)$ for values of $(\Mx, \My)$ in the square of side two.  Each
point in the square is mapped to a set of conditional
probabilities $p (\lambda_i \vert u, v)$.  The yellow horizontal hatched, red vertical hatched, and
blue diagonal hatched regions, with $\Mx \geq \My$, are mapped according to the
entries in Tables~\ref{tab:general2}--\ref{tab:general4}.  Tables
corresponding to the green horizontal hatched, orange vertical hatched, and purple diagonal hatched regions, with $\My
\geq \Mx$, may be constructed by swapping Alice's and Bob's
detector-setting labels, $x \leftrightarrow y, x' \leftrightarrow
y'$ as well as the labels $1 \leftrightarrow 2$.  The model
saturates the general \two-parameter bound, providing an
additional proof that the bound is tight.
}
\label{fig:freedom_square2}
\end{figure}

The interpolating model is deterministic and locally causal, with a hidden variable $\lambda$ that can take on one of $5$ discrete values, $\lambda_1, \lambda_2, \dots, \lambda_5$. For this model the deterministic measurement-outcome functions $A (u , \lambda_i), B (v , \lambda_i)$ for Alice and Bob are of the forms defined in Table~\ref{tab:general1}, where the constants $c, d, e, f, g$ may be any values in  $\{-1, 1\}$. We divide the square of possible $(\Mx,\My)$ values into six regions, as shown in Fig.~\ref{fig:freedom_square2}, and for each region we construct a mapping from the parameters $(\Mx,\My)$ to a set of conditional probabilities, using Tables~\ref{tab:general2}-\ref{tab:general4}, as follows.

%%%%%%%%%%%%%%%%%%%%%%%%%%%%%%%%%%%%%%%%
%%%%%%%%%%%%%%%%%%%%%%%%%%%%%%%%%%%%%%%%

Consider first the yellow horizontal hatched region in
Fig.~\ref{fig:freedom_square2}, which corresponds to $\Mx \geq
\My$, $0 \leq \Mx \leq 2$, $0 \leq \My \leq 2/3$, and $\Mx + 2
\My \leq 2$.  For this region, the mapping of the interpolating
model will be defined by Table~\ref{tab:general2}.  Using
Eq.~(\ref{abexpuvM}) and Tables \ref{tab:general1} and
\ref{tab:general2}, we find that the CHSH parameter $S$ of
Eq.~(\ref{bellE}) takes the form
%%%%%%%
\begin{equation}
S = 2 + 2 \pxone + 4 \pxtwo \, .
\label{Stable}
\end{equation}
Assuming that $\pxone \ge \pxtwo$, the degrees of measurement
dependence $\Mx$ and $\My$, as defined by
Eqs.~(\ref{hallMaCHSH1})-(\ref{hallMaCHSH2}), are found from
Table~\ref{tab:general2} to be
%%%%%%
\begin{equation}
\begin{aligned}
    \Mx &=& {\rm max} \{ 2\pxone , 2\pxone \} = 2 \pxone , \\
    \My &=& {\rm max} \{ 2\pxtwo , 2 \pxtwo \} = 2 \pxtwo .
\label{Mtabs12}
\end{aligned}
\end{equation}
(For Table~\ref{tab:general2}, we also find that $M = {\rm max}
\{ \Mx, \My \} = \Mx$.) Thus, Table~\ref{tab:general2} with
$\pxone \ge \pxtwo$ is a potential model for the region of
parameter space for which $\Mx \ge \My$, with $p_1 = \Mx/2$ and
$p_2 = \My/2$.  To be a viable model, the conditional
probabilities that it defines must be nonnegative.  This will be
the case provided that $1 - \pxone - 2 \pxtwo = 1 - \frac{\Mx}{2}
- \My \ge 0$, so the region of validity is precisely the yellow horizontal hatched
region of Fig.~\ref{fig:freedom_square2}.  (The total probability
for each setting pair must sum to unity, but this can be seen
immediately by summing each column of Table~\ref{tab:general2}.)

Given that $p_1 = \Mx/2$ and $p_2 = \My/2$, Eq.~\eqref{Stable}
implies that
\begin{equation}
S = 2 + \Mx + \My + {\rm min} \{ \Mx, \My \} ,
\label{StableM}
\end{equation}
saturating the upper bound derived in Eq.~(\ref{GeneralS}) for
the case in which $\Mx + \My + {\rm min} \{ \Mx, \My \} \leq 2$.

We proceed similarly for the red vertical hatched and blue diagonal hatched regions of
Fig.~\ref{fig:freedom_square2}, which both satisfy $\Mx \ge \My$
and $\Mx + 2 \My \geq 2$.  The red vertical hatched region corresponds to $2/3
\leq \Mx \leq 2$ and $0 \leq \My \leq 2/3$, and is described in
the interpolating model by Table~\ref{tab:general3}\null.  Assuming
that $\pxone \le 1$, $\pxtwo \le 1/3$, and $\pxone + 2 \pxtwo \ge
1$, we find from Table~\ref{tab:general3} that
\begin{equation}
\begin{aligned}
    \Mx &=&& {\rm max} \{ 2\pxone , 1 + \pxone - 2 \pxtwo\} &&= 2 \pxone , \\
    \My &=&& {\rm max} \{ 2\pxtwo , 1- \pxone \} &&= 2 \pxtwo .
\label{Mtabs12b}
\end{aligned}
\end{equation}
(We again find that $M = \Mx$.)  In this case no additional
constraints are imposed by nonnegativity, but the constraints that
we imposed to evaluate $\Mx$ and $\My$ are precisely the
conditions that delineate the red vertical hatched region of
Fig.~\ref{fig:freedom_square2}.

The blue diagonal hatched region corresponds to $2/3 \leq \Mx \leq 2$, $2/3 \leq
\My \leq 2$, again with $\Mx \geq \My$, and is described in the
interpolating model by Table~\ref{tab:general4}, with
\begin{equation}
\begin{aligned}
  \pxone &= \frac{2 - \My}{4} + \frac{\Mx-\My}{12} \, ,\\
  \pxtwo &= \frac{\Mx - \My}{6} \, ,\\
  \pxthree &= \begin{cases}
      0 & \text{if } \Mx \le 4 \My - 2 \, ,\\
      \frac{1}{8} (\Mx - 4 \My + 2) & \text{otherwise.}
  \end{cases}
  \label{Mtabs12c}
\end{aligned}
\end{equation}
When $\Mx = \My$, $\pxtwo$ and $\pxthree$ vanish, and $\pxone$
becomes equal to the value of $p$ in Table II of Hall
\cite{hall10}, making our Table~\ref{tab:general4} an
exact match for Table II of Hall.  Using Table~\ref{tab:general4}
and Eqs.~\eqref{Mtabs12c}, one can verify that
\begin{equation}
\begin{aligned}
   \Mx[\y] &= \Mx \, ,\\
   \Mx[\yp] &= \begin{cases}
      2 \My - \Mx & \text{if } \Mx \le 4 \My - 2 \, ,\\
      2 - 2 \My & \text{otherwise,}
   \end{cases} \\
   \My[\x] &= \My \,,  \\
   \My[\xp] &= \My \, ,\\
\end{aligned}
\end{equation}
where we are using the definitions in Eqs.~\eqref{Mx[v]} and
\eqref{My[u]}.  For the parameter range of the blue diagonal hatched region, it is
easily seen that $\max(\Mx[\y],\Mx[y']) = \Mx$, as desired.  One
can also verify that the probabilities in each column of
Table~\ref{tab:general4} sum to 1, and that all the entries of
the table are nonnegative for $(\Mx,\My)$ in the blue diagonal hatched region of
Fig.~\ref{fig:freedom_square2}.  Thus, in the blue diagonal hatched region
Table~\ref{tab:general4} defines a consistent set of conditional
probabilities that match the Hall model when $\Mx = \My$.  (We
again find that $M = \Mx$.)

Upon using Tables \ref{tab:general3} and \ref{tab:general4} together with the measurement outcomes in Table \ref{tab:general1}, we find $S = 4$ for both the red vertical hatched and blue diagonal hatched regions, saturating the upper bound in Eqs.~(\ref{GeneralS})-(\ref{bound1}). To our knowledge, the model represented by Tables \ref{tab:general1}--\ref{tab:general4} has not been described previously in the literature.

Thus the interpolating model, as defined by Tables
\ref{tab:general1}--\ref{tab:general4}, saturates the
\two-parameter general bound for all values $\Mx \geq \My$.  By
symmetry, one can complete the definition of the interpolating
model by constructing equivalent tables for $\My \geq \Mx$, by
switching settings labels $x \leftrightarrow y$, $x'
\leftrightarrow y'$ and subscripts $1 \leftrightarrow 2$.  Since
the interpolating model saturates the two-parameter bound of
Eq.~\eqref{bound1}, it provides an additional proof that
Eq.~(\ref{bound1}) is a tight upper bound on $S$ for
hidden-variable models that obey local causality but do not obey
measurement independence. 

To show how the interpolating model is related to the Hall model
of Table I of Ref.~\cite{hall10}, and the Banik \etal model of
Table 1 of Ref.~\cite{banik12}, we introduce a notation that uses
subscripts to show explicitly the dependence of the conditional
probabilities $p(\lambda_i|u,v)$ on the parameters $\pxone$ and
$\pxtwo$.  In particular, we will denote the entries of
Table~\ref{tab:general2} by
\begin{equation}
p^{\ref{tab:general2}}_{\pxone,\pxtwo} (\lambda_i|u,v) \, ,
\end{equation}
and the entries of the Hall model by
\begin{equation}
p^{H}_{p} (\lambda_i|u,v) \, .
\end{equation}
Table I of Ref.~\cite{banik12} has only two rows, but they can be
identified with rows 3 and 5 of the other models, with the
remaining rows set to zero.  Thus, the conditional probabilities
of the Banik \etal model can be denoted by
\begin{equation}
p^{B}_{p} (\lambda_i|u,v) \, .
\end{equation}
It is then easily seen that for $\Mx \ge \My$ and $\Mx + 2 \My
\le 2$, when Table~\ref{tab:general2} applies, the interpolating
model exactly matches the two previous models in the appropriate
limits:
\begin{align}
p^{\ref{tab:general2}}_{p,p} (\lambda_i|u,v) &= p^{H}_p
(\lambda_i|u,v) \ ,
\label{HallLim}\\
p^{\ref{tab:general2}}_{p,0} (\lambda_i|u,v) &= p^{B}_p
(\lambda_i|u,v) \ .
\label{BanikLim}
\end{align}
Furthermore, for general values it is simply a linear
interpolation:
\begin{equation}
p^{\ref{tab:general2}}_{\pxone,\pxtwo} (\lambda_i|u,v) = w
p^{H}_{\pxone} (\lambda_i|u,v) + (1 - w) p^{B}_{\pone}
(\lambda_i|u,v) \, ,
\end{equation}
where $w = \pxtwo/\pxone$.

%%%%%%%%%%%%%%%%%%%%%%%%%%%%%%%%%%%%%%%%
%TABLE I
%%%%%%%%%%%%%%%%%%%%%%%%%%%%%%%%%%%%%%%%
\linespread{1.0}
\begin{table}[t]
\centering
\begin{tabular}{| c | c| c| c| c | } 
\toprule[1.5pt]
$\lam_i$       & $\Xlam$ & $\Xlamp$ & $\Ylam$ & $\Ylamp$ \\
\toprule[1.5pt]
${\lam}_1$ & $c$ & $c$ & $c$ & $c$ \\
${\lam}_2$ & $d$ & $-d$ & $d$ & $d$ \\
${\lam}_3$ & $e$ & $e$ & $e$ & $-e$ \\
${\lam}_4$ & $f$ & $-f$ & $-f$ & $f$ \\
${\lam}_5$ & $g$ & $g$ & $g$ & $g$ \\
\bottomrule[1.25pt]
\end{tabular}\par
\caption{
Deterministic measurement-outcome functions $A (u , \lambda_i)$ and $B (v , \lambda_i)$ for Alice's and Bob's measurements, given $\lambda_i$ with $i = 1, ... , 5$. The values of the measurement outcomes ($c, d, e, f, g$) are selected  arbitrarily from $\{ -1, 1 \}$.
\label{tab:general1}
}
\end{table}
%%%%%%%%%%%%%%%%%%%%%%%%%%%%%%%%%%%%%%%%

%%%%%%%%%%%%%%%%%%%%%%%%%%%%%%%%%%%%%%%%
%TABLE II
%%%%%%%%%%%%%%%%%%%%%%%%%%%%%%%%%%%%%%%%
\linespread{1.0}
\begin{table}[t]
\centering
\begin{tabular}{| c | c | c | c | c |} 
\toprule[1.5pt]
$\lam_i$        & $\pxyl$        & $\pxypl$ & $\pxpyl$ & $\pxpypl$ \\
\toprule[1.5pt]
${\lam}_1$   & $\pxtwo$                      & $\pxtwo$                   & $ \pxtwo$                   & $0$ \\
${\lam}_2$  & $\pxtwo$                       & $\pxtwo$                    & $0$                             & $\pxtwo$ \\
${\lam}_3$  & $\pxtwo$                       & $0$                             & $\pxone$                    & $\pxone$ \\
${\lam}_4$ & $0$                                & $\pxtwo$                    & $\pxtwo$                    & $\pxtwo$ \\
${\lam}_5$  &  $\ 1-3\pxtwo \ $ & $1-3\pxtwo \ $ & $1-\pxone - 2\pxtwo \ $ & $1-\pxone - 2\pxtwo$ \\
\bottomrule[1.25pt]
\end{tabular}\par
\caption{
Conditional probabilities $p (\lambda_i \vert u,v)$ for the value of the hidden variable $\lambda$ to be $\lambda_i$, for $\Mx + \My + {\rm min} \{ \Mx, \My \} \leq 2$, $0 \leq \Mx \leq 2$, $0 \leq \My \leq 2/3$, and $\Mx \geq \My$ (Fig.~\ref{fig:freedom_square2} yellow horizontal hatched region).
\label{tab:general2}
}
\end{table}
%%%%%%%%%%%%%%%%%%%%%%%%%%%%%%%%%%%%%%%%

%%%%%%%%%%%%%%%%%%%%%%%%%%%%%%%%%%%%%%%%
%TABLE III
%%%%%%%%%%%%%%%%%%%%%%%%%%%%%%%%%%%%%%%%
\linespread{1.0}
\begin{table}[ht]
\centering
\begin{tabular}{| c | c | c | c | c |} 
\toprule[1.5pt]
$\lam_i$    & $\pxyl$           & $\pxypl$          & $\pxpyl$              & $\pxpypl$ \\
\toprule[1.5pt]
${\lam}_1$  & $\pxtwo$          & $\pxtwo$          & $\frac{1}{2} (1-\pxone)$  & $0$                             \\
${\lam}_2$  & $\pxtwo$          & $\pxtwo$          & $0$                   & $\frac{1}{2} (1-\pxone)$                    \\
${\lam}_3$  & $\pxtwo$          & $0$               & $\frac{1}{2} (1-\pxone)$        & $\frac{1}{2} (1-\pxone)$ \\
${\lam}_4$  & $0$               & $\pxtwo$          & $\pxone$              & $\pxone$                                  \\
${\lam}_5$  &  $\ 1-3\pxtwo \ $ & $1-3\pxtwo \ $    & $0$                   & $0$ \\
\bottomrule[1.25pt]
\end{tabular}\par
\caption{
Conditional probabilities $p (\lambda_i \vert u,v)$ for the case $\Mx + \My + {\rm min} \{ \Mx, \My \} \geq 2$, $2/3 \leq \Mx \leq 2$, $0 \leq \My \leq 2/3$, and $\Mx \geq \My$ (Fig.~\ref{fig:freedom_square2} red vertical hatched region).
\label{tab:general3}
}
\end{table}
%%%%%%%%%%%%%%%%%%%%%%%%%%%%%%%%%%%%%%%%

%%%%%%%%%%%%%%%%%%%%%%%%%%%%%%%%%%%%%%%%
%TABLE IV
%%%%%%%%%%%%%%%%%%%%%%%%%%%%%%%%%%%%%%%%
\def\ss{\scriptstyle}
\linespread{1.0}
\begin{table}[ht]
\centering
\begin{tabular}{| c | c | c | c | c |} 
\toprule[1.5pt]
$\lam_i$    & $\pxyl$               & $\pxypl$              & $\pxpyl$              & $\pxpypl$ \\
\toprule[1.5pt]
${\lam}_1$  & $\ss\pxone-2\pxtwo$ & $\frac{1-\pxone}{2}\ss-2\pxtwo+\pxthree$ 
     & $\frac{1-\pxone}{2}\ss+\pxtwo-\pxthree$ & $\ss 0$ \\
${\lam}_2$  & $\frac{1-\pxone}{2}\ss+4\pxtwo-\pxthree$ & $\ss\pxone+\pxtwo$ 
     & $\ss 0$ & $\frac{1-\pxone}{2}\ss+\pxtwo - \pxthree$\\
${\lam}_3$  & $ \frac{1-\pxone}{2}\ss-2\pxtwo+\pxthree$ & $\ss 0$ 
     & $\ss\pxone-2\pxtwo +2\pxthree$ 
     & $\frac{1-\pxone}{2}\ss-2\pxtwo+3\pxthree$ \\
${\lam}_4$  & $\ss 0$ & $\frac{1-\pxone}{2}\ss+\pxtwo - \pxthree$ 
     & $\frac{1-\pxone}{2}\ss+\pxtwo - \pxthree$ 
     & $\ss\pxone+\pxtwo-2\pxthree$ \\
${\lam}_5$  & $\ss 0$ & $\ss 0$ & $\ss 0$ & $\ss 0$ \\
\bottomrule[1.25pt]
%%%%%%%%%%%%%%
\end{tabular}\par
\caption{
Conditional probabilities $p (\lambda_i \vert u,v)$ for the case $\Mx + \My + {\rm min} \{ \Mx, \My \} \geq 2$, $2/3 \leq \Mx, \My \leq 2$, and $\Mx \geq \My$ (Fig.~\ref{fig:freedom_square2} blue diagonal hatched region).
\label{tab:general4}
}
\end{table}
%%%%%%%%%%%%%%%%%%%%%%%%%%%%%%%%%%%%%%%%

%-----------------------------------------------------------------------------------------------------------------------------------------------
 \section{Mutual Information for the Interpolating Model}
 \label{sec:info1}
 %-----------------------------------------------------------------------------------------------------------------------------------------------
 
 Just as in Sec.~\ref{sec:info}, we can compute the mutual information between the hidden variable $\lambda$ and the
detector settings for
the interpolating model, using
Eqs.~(\ref{eq:mutualinf})-(\ref{HpLambdaUV}) and the
measurement outcomes from Table~\ref{tab:general1}. Here
we consider the case $\Mx \geq
\My$ and the range $\Mx + 2 \My \leq 2$ with violations $V\in
[0,2]$, so Table~\ref{tab:general2} applies.

\begin{figure}
\centering
\begin{tabular}{@{}c@{}}
\includegraphics[width=3.25in]{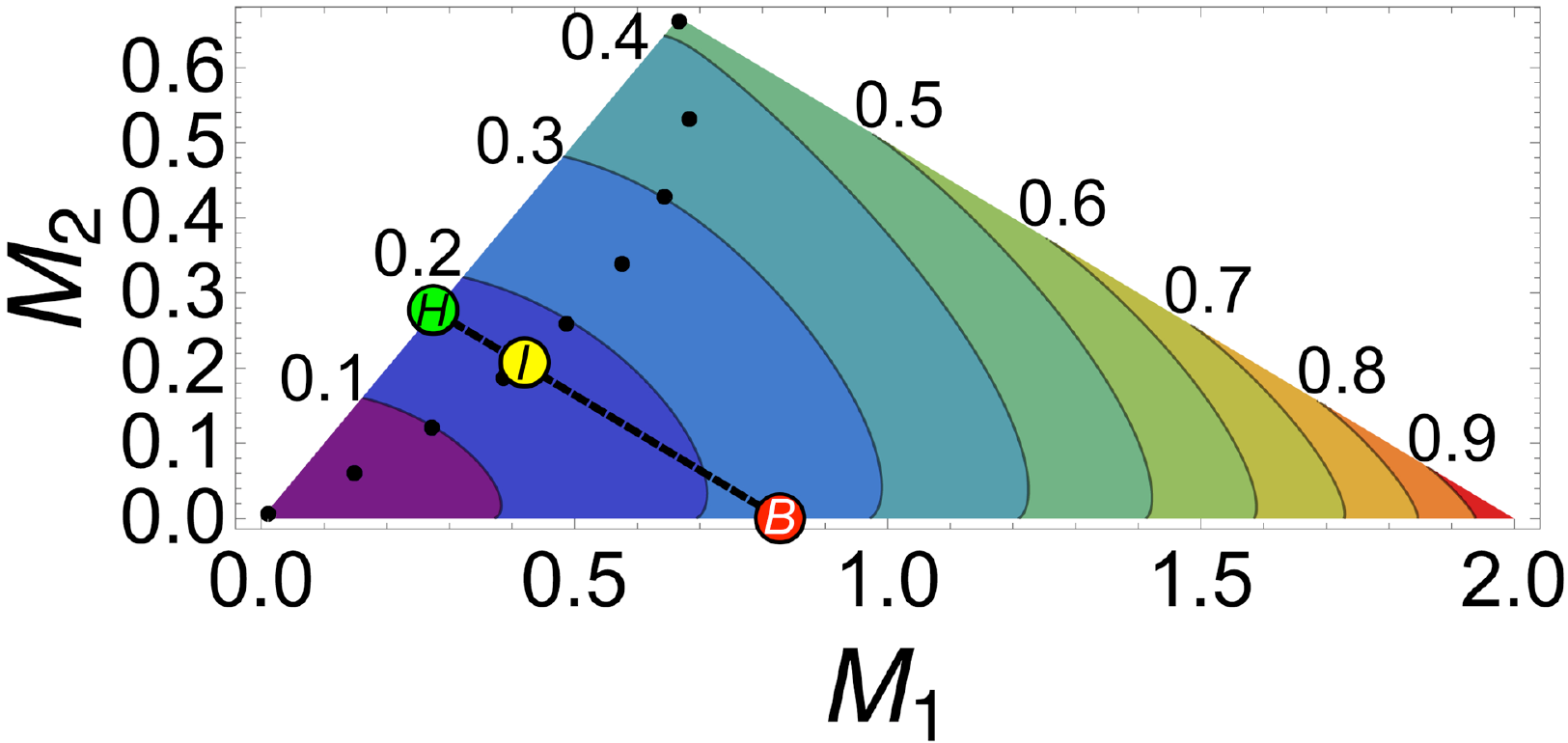} 
\end{tabular}
\caption{
\baselineskip 9pt 
``Freedom Square" plot 
%color coded 
labeled by contours of 
mutual information $I_I(\Mx,\My)$ for the interpolating model, in bits, from Eq.~(\ref{IGM1M2}) derived from Table~\ref{tab:general2}. The Hall and Banik \etal subcases are denoted by light green ($H$) and dark red ($B$) circles, respectively, with the light yellow circle ($I$) showing where mutual information is minimized for the interpolating model. The dashed line is the slice satisfying $V(\Mx,\My)=\Vs$, connecting solutions for the Hall model $(\Mx,\My)=(\Vs/3,\Vs/3)$, and the Banik model $(\Mx,\My)=(\Vs,0)$, with minimum mutual information $I_I(\bar \Mx, \bar \My) \approx 0.1616$ at $(\bar\Mx,\bar\My) \approx (0.4158,0.2063)$. See Fig.~\ref{fig:freedom_square3a} (top). 
The black dots, plus the light yellow circle ($I$), trace the curve 
that minimizes the mutual information for each value of Bell violation $V \in [0,2]$.
See Fig.~\ref{fig:freedom_square3a} (bottom).
}
\label{fig:freedom_square3}
\end{figure}

Using Table~\ref{tab:general2} and Eq.~(\ref{IGa}), and recalling that $p_1=\Mx/2$, $p_2=\My/2$ in this model, we find that the mutual information for the interpolating model is given by
%\begin{eqnarray}
%&I_I&(\Mx,\My)   =  \frac{1}{4}\sum_{i,u,v}p(\lambda_i|u,v)\log_2 p(\lambda_i|u,v) \nonumber\\ 
%& &\qquad\qquad\qquad\qquad 
% -\sum_i p(\lambda_i)\log_2 p(\lambda_i)  \nonumber \\
%& = & \frac{1}{4}\Bigg\{   \Mx \log_2\left[\frac{\Mx}{2}\right] +  \frac{\My}{2}\left( %\log_2\left[ \frac{\My}{2} \right]+9\log_2\frac{4}{3}\right)  \nonumber\\
%&  & \qquad + (2-3\My) \log_2\left[ \frac{2-3\My}{2}\right]  \nonumber \\ 
%&  &\qquad +   (2 - \Mx - 2\My)\log_2\left[ \frac{2-\Mx-2\My}{2} \right] \nonumber \\ 
%&  & \qquad - \frac{ (2\Mx +\My) }{2} \log_2\left[\frac{2\Mx +\My}{8} \right]  \nonumber \\  
%& & \qquad - (4-\Mx-5\My) \log_2\left[ \frac{4-\Mx-5\My}{4}  \right]  \Bigg\}.
%\label{IGM1M2}
%\end{eqnarray}

\begin{eqnarray}
&I_I&(\Mx,\My)   =  \frac{1}{4}\sum_{i,u,v}p(\lambda_i|u,v)\log_2 p(\lambda_i|u,v) \nonumber\\ 
& &\qquad\qquad\qquad\qquad 
 -\sum_i p(\lambda_i)\log_2 p(\lambda_i)  \nonumber \\
& = & \frac{1}{4}\Bigg\{  2h\left(\frac{2-3\My}{2}\right) + 2h\left(\frac{2-\Mx-2\My}{2}\right)  \nonumber\\
&  & \qquad - 4h\left(\frac{2\Mx + \My}{8}\right) - 4h\left(\frac{4-\Mx-5\My}{4}\right) \nonumber \\
&  & \qquad + 2h\left(\frac{\Mx}{2}\right) + h\left(\frac{\My}{2}\right) + \frac{9\My}{2} \log_2\frac{4}{3} \Bigg\} 
\label{IGM1M2}
\end{eqnarray}
Eq.~(\ref{IGM1M2}) is plotted in Fig.~\ref{fig:freedom_square3}.
%\andy{Simplify with $h(x)$ notation.}
 
%%%%%%%%%%%%%%%%%%%%%%%%%%%%%%%%%%%%%%%%
%%%%%%%%%%%%%%%%%%%%%%%%%%%%%%%%%%%%%%%%
 
%%%%%%%%%%%%%%%%%%%%%%%%%%%%%%%%%%%%%%%%
\begin{figure}
\centering
\begin{tabular}{@{}c@{}}
\includegraphics[width=3.3in]{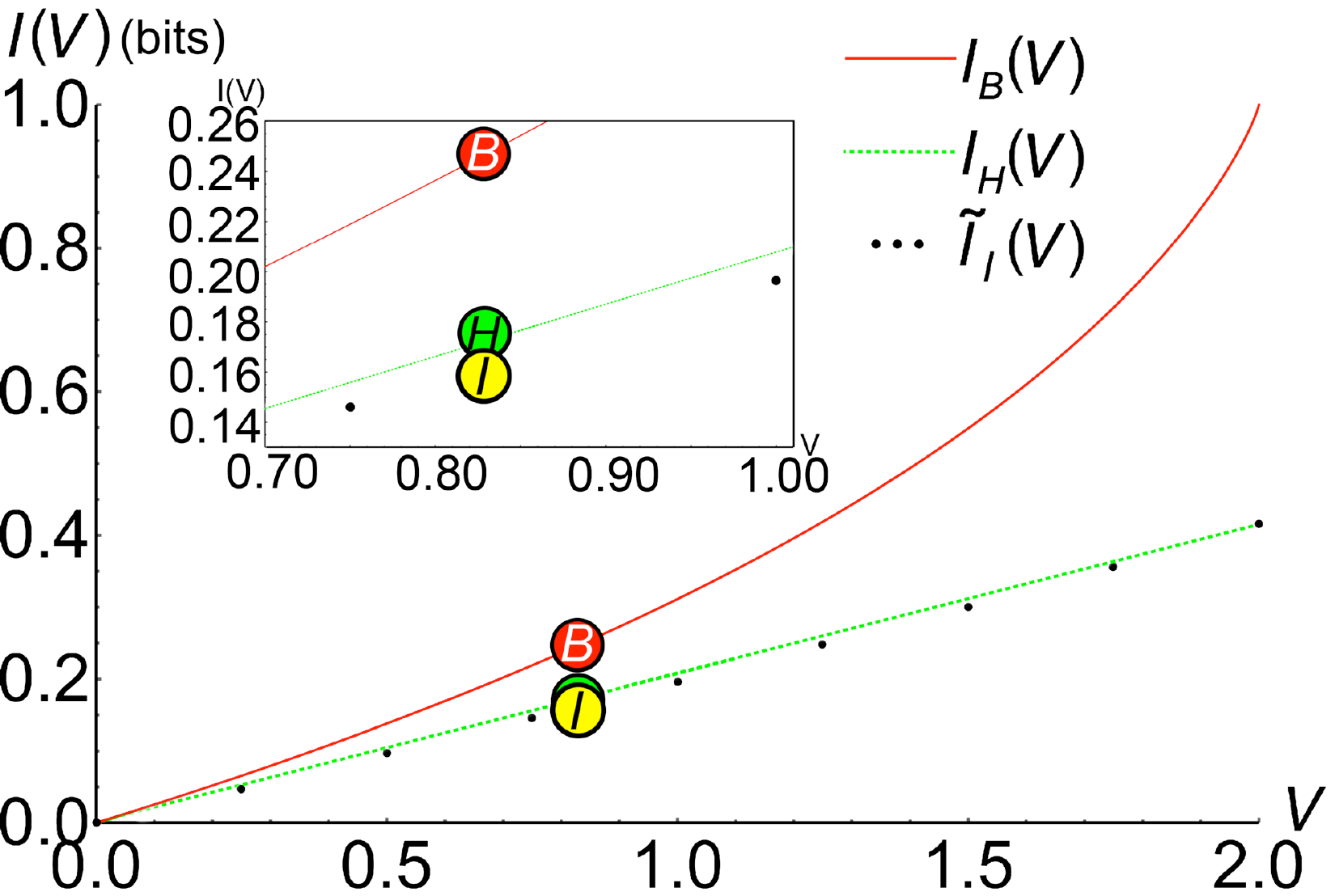} \\
\includegraphics[width=3.3in]{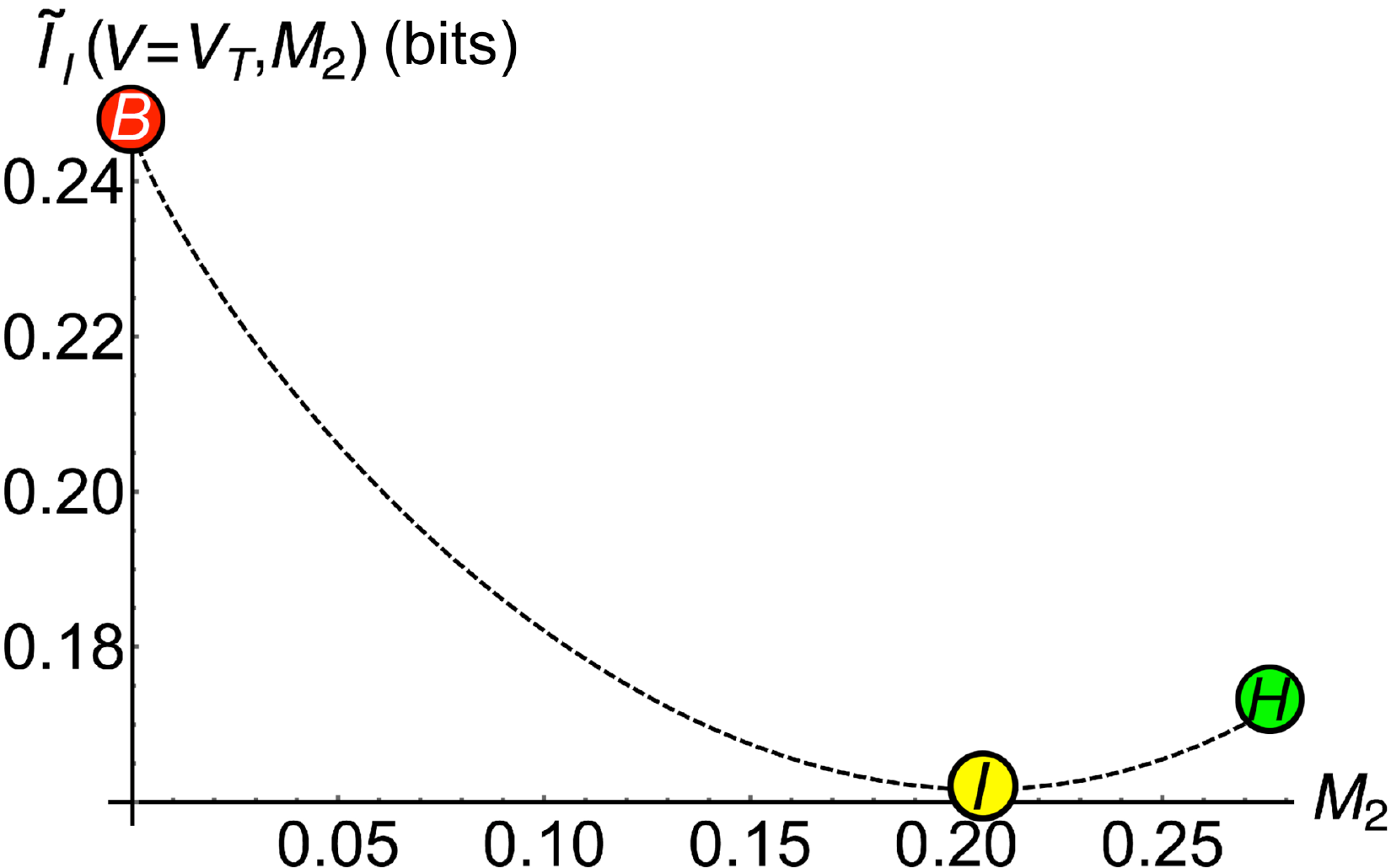} \\
\end{tabular}
\caption{
\baselineskip 9pt 
These plots use the 
interpolating model  
with $V(\Mx,\My) = \Mx + 2\My \leq 2$, $\Mx \geq \My$. 
As in Fig.~\ref{fig:freedom_square3}, large colored circles marked with $B$, $H$ or $I$ denote $V(\Mx,\My)=\Vs$ for the Banik, Hall, and interpolating models, respectively. The interpolating model, which for these parameters is defined by Table~\ref{tab:general2}, requires less mutual information (light yellow circle) to produce a given Bell violation than the previously studied Hall and Banik \etal subcases, denoted by light green and darkred circles, respectively. ({\it Top}) Mutual information for the Hall model (dotted green curve), the Banik {\it et al.} model (solid red curve), and the minimum of the interpolating model $\tilde{I}_I(V) \equiv \min_{\My} \tilde{I}_I(V,\My)$
(black dots, as in Fig~\ref{fig:freedom_square3}), 
in bits, plotted as a function of CHSH violation $V\in[0,2]$ (see inset plot).
The Hall model always requires less mutual information than the Banik model to produce a given Bell violation, while
the minimum of the interpolating model requires even less mutual information than the Hall or Banik models. ({\it Bottom}) Mutual information 
$\tilde{I}_I(\Vs,\My)$ required to reach the Tsirelson bound $V(\Mx,\My)=\Vs$ (e.g. dashed black line in Fig.~\ref{fig:freedom_square3}), plotted in bits as a function of  $\My\in[0,\Vs/3]$. 
}
\label{fig:freedom_square3a}
\end{figure}
%%%%%%%%%%%%%%%%%%%%%%%%%%%%%%%%%%%%%%%%

Using the relation $\Mx = V - 2\My$ for models that saturate the general \two-parameter bound, we can also express $I_I$ in terms of $\My$ and the amount of violation $V$:
%\begin{eqnarray}
%&\tilde{I}_{I}&(V,\My)  \equiv I_I(V-2M_2,M_2) \nonumber \\ 
%&=& \frac{1}{4}\Bigg\{ (V-2\My) \log_2\left[\frac{V-2\My}{2}\right] \nonumber\\
%&  &\qquad + \frac{\My}{2}\left( \log_2\left[ \frac{\My}{2} \right]+9\log_2\frac{4}{3}\right)  \nonumber\\
%& & \qquad + (2-3\My)\log_2\left[ \frac{2-3\My}{2}\right] \nonumber \\ 
%& & \qquad + (2-V) \log_2\left[ \frac{2-V}{2} \right] \nonumber \\ 
%&  & \qquad - \frac{ ( 2V - 3\My) }{2} \log_2\left[\frac{2V -3\My}{8} \right]    \nonumber \\ 
%&  & \qquad - (4-V-3\My) \log_2\left[ \frac{4-V-3\My}{4}  \right]  \Bigg\} ,
%\label{IGM1M2V}
%\end{eqnarray} 
\begin{eqnarray}
&\tilde{I}_{I}&(V,\My)  \equiv I_I(V-2M_2,M_2) \nonumber \\ 
&=& \frac{1}{4}\Bigg\{ 2 h\left(\frac{V-2\My}{2}\right)  + h\left(\frac{\My}{2}\right) + \frac{9\My}{2} \log_2\frac{4}{3}  \nonumber\\
& & \qquad - 4h\left(\frac{2V-3\My}{8}\right) - 4h\left(\frac{4-V-3\My}{4}\right)  \nonumber \\ 
&  & \qquad + 2h\left(\frac{2-3\My}{2}\right) + 2h\left(\frac{2-V}{2}\right) \Bigg\} ,
\label{IGM1M2V}
\end{eqnarray} 
with $\My$ restricted to the range $0\leq \My\leq V/2\leq 2$. 
We denote the minimum of $\tilde{I}_{I}(V,\My)$,
minimized over $\My$, by $\tilde{I}_{I}(V)$.
%\andy{Simplify with $h(x)$ notation.}

The mutual information requirements of the Hall model of
Ref.~\cite{hall10} and the Banik \etal model of
Ref.~\cite{banik12} were discussed in Sec.~\ref{sec:info}.  The
mutual information required for each model, to achieve a
specified Bell--CHSH inequality violation $V$, was specified in 
Eq.~\eqref{eq:Hallmutual1} for $I_H(V)$ and
Eq.~\eqref{eq:Banikmutual1} for $I_B(V)$.  These functions were
plotted in Fig.~\ref{fig:freedom_square3d} in comparison with
$\tilde{I}_G(V)$, the minimum mutual information for the
\two-parameter model of Sec.~\ref{sec:model}.  In the top panel
of Fig.~\ref{fig:freedom_square3a}, the same two functions are
shown in comparison with $\tilde {I}_I(V)$, the minimum mutual
information for the interpolating model.  $\tilde {I}_I(V)$ is
less than either of these two comparison models, but it is
nonetheless significantly larger that the mutual information
required by the \two-parameter model of Sec.~\ref{sec:model}. 
For the maximum quantum violation $\Vs$, the Banik \etal model
requires 0.247 bits of mutual information, the Hall model
requires 0.172 bits, and the interpolating model requires 0.162
bits.  The \two-parameter model of Sec.~\ref{sec:model} requires
only 0.0462 bits, as shown in Eq.~\eqref{eq:IGmin}.

The lower panel of Fig.~\ref{fig:freedom_square3a} shows the
mutual information of the interpolating model,
$\tilde{I}_I(\Vs,\My)$, for the maximum quantum violation $\Vs$,
shown in bits as a function of $\My$.  The minimum occurs at $\Mx
\approx 0.416$ and $\My \approx 0.206$ (yellow circle ($I$)).

Overall, while the interpolating model requires less mutual information between the settings and hidden variables to mimic the quantum predictions for violations of the Bell-CHSH inequality than previously studied locally causal models, it is significantly less efficient than the \two-parameter model of Sec.~\ref{sec:model}.

\section{Steps in Proof of the Four-Parameter Relaxed Bell-CHSH Inequality}
\label{sec:4boundproof}

We wish to prove that the inequality 
\begin{equation}
T_1 \le 2 + \Mx[y'] \, .
\label{T1-4p2appendix}
\end{equation}
from Eq.~\eqref{T1-4p2} holds.

Starting with
Eq.~\eqref{bellE1}, we replace Eq.~\eqref{bellE2} with
\begin{equation}
\begin{aligned}
\ES  &=  \Bigg| \int d\lam \Big\{\pxypl  \Big[ \Xlam \Ylamp  + \Xlam \Ylam \Big]   \\
  & \quad -  \pxpypl \Big[ \Xlamp \Ylamp - \Xlamp \Ylam \Big]  \\
  & \quad + \Xlam \Ylam \Big[ \pxyl - \pxypl \Big]  \\
  & \quad + \Xlamp \Ylam \Big[ \pxpyl - \pxpypl \Big] \Big\} \Bigg| .
 \end{aligned}
\label{bell4p-E2}
\end{equation}
Then
\begin{equation}
\ES  \leq \tilde T_1 + \tilde T_2 + \tilde T_3 \, ,
\label{bell4p-E3}
\end{equation}
where
\begin{equation}
\begin{split}
\tilde T_1 &= \int d\lam \Big|  \pxypl  \Big[ \Xlam \Ylamp  + \Xlam \Ylam \Big]    \\
  &\quad\quad - \pxpypl \Big[ \Xlamp \Ylamp - \Xlamp \Ylam \Big] \Big|,
\end{split}
\label{bell4p-T1}
\end{equation}
\begin{equation}
\tilde T_2 = \int d\lam \Big| \Xlam \Ylam \Big[ \pxyl - \pxypl \Big] \Big|,
\label{bell4p-T2}
\end{equation}
and
\begin{equation}
\tilde T_3 =  \int d\lam \Big| \Xlamp \Ylam \Big[ \pxpyl - \pxpypl \Big] \Big|.
\label{bell4p-T3}
\end{equation}
Clearly 
\begin{equation}
\tilde T_2 + \tilde T_3 = \My + \Mty \, ,
\end{equation}
and $\tilde T_1$ can be rewritten as 
\begin{widetext}
\begin{eqnarray}
\tilde T_1 &=& \int d\lam \Bigg|  \Ylam \Big[ \Xlam \pxypl  + \Xlamp \pxpypl \Big] 
 \nonumber 
   +  \Ylamp \Big[ \Xlam \pxypl - \Xlamp \pxpypl \Big] \Bigg|
 \label{bell4p-T1s} \\
&\leq &  \int d\lam \Bigg\{ \Bigg|  \Ylam \Xlam \bigg[ \pxypl  + \frac{\Xlamp}{\Xlam} \pxpypl \bigg] \Bigg| \nonumber 
  +  \Bigg| \Ylamp \Xlam \bigg[ \pxypl - \frac{\Xlamp}{\Xlam} \pxpypl \bigg] \Bigg| \Bigg\}
 \label{bell4p-T1s1} \\
 &\leq&  \int d\lam \Bigg\{ \Bigg| \pxypl  + \frac{\Xlamp}{\Xlam} \pxpypl  \Bigg|  \nonumber 
   + \Bigg| \pxypl - \frac{\Xlamp}{\Xlam} \pxpypl \Bigg|   \Bigg\} \, . 
\label{bell4pT1s2} \\
&\leq& 2 + \Mx[y'] \ ,
\label{bell4p-T1s3}
\end{eqnarray}
\end{widetext}
as we had claimed.

\section{Construction of the Four-Parameter Model}
\label{sec:4pmodel}

The \two-parameter model of Sec.~\ref{sec:model} was found by
trial and error, but attempts at finding a \four-parameter model
using trial and error did not succeed.  But there is a more
systematic way, based on examining the proof of the bound in
Sec.~\ref{sec:4pbound}, identifying the conditions that are needed
to saturate it. We describe this systematic approach in some detail in this Appendix, as we believe the basic ideas are of general value to the construction of saturating models. 

Without loss of generality we can seek a solution with $\Mx \ge
\My$, because the opposite case can be treated by interchanging
the labels 1 and 2, which is equivalent to interchanging
$(\x,\xp)$ with $(\y,\yp)$.  Similarly, we can without loss of
generality seek a solution with $\Mx[\y] \ge \Mx[\yp]$, because
the opposite case can be treated by interchanging the labels $\y$
and $\yp$.  Thus, our solution will have $\Mx = \Mx[\y]$, and
$\Mtx = \Mx[\yp]$.  Finally, we can without loss of generality
seek a solution with $\My[\x] \ge \My[\xp]$, so the solution will
have $\My = \My[\x]$, and $\Mty = \My[\xp]$.

The proof involved showing two bounds on $T_1$: $T_1 \le 2 +
\Mx[y]$ and $T_1 \le 2 + \Mx[\yp]$.  For the conventions adopted
in the previous paragraph, it is the second of these bounds that
is the more stringent, so it is the second that must be
saturated.  This means that we must examine the bound that was
demonstrated in Appendix~\ref{sec:4boundproof}, Eqs.~\eqref{bell4p-E2}--\eqref{bell4pT1s2}.  We
initially restrict ourselves to the case $\My + \Mtx + \Mty \le
2$, since it is only in this case that the bound shown in
Eq.~\eqref{bell4p-T1s6} is saturated.

Starting with Eq.~\eqref{bell4p-E2} for $\ES$, we recognize that
the integral over $\lambda$ reduces for our model to the sum over
the four values of $\lambda$: $\lambda_1 \ldots \lambda_4$, as
listed in Table~\ref{tab:general1a}.  The bound is established by
rewriting the integrand as the sum of judiciously chosen pieces,
and then bounding the absolute value of the integral by the sum
of the integrals of the absolute values of the pieces.  The bound
will therefore be equal to $\ES$ if each of the pieces is
positive, so the absolute value signs become irrelevant.  (The
bound would also be saturated if all the pieces were negative,
but we did not pursue this option.)  Thus, we will examine each
piece, and insist that it be positive.

We start with the third line of Eq.~\eqref{bell4p-E2}, which are the
terms that are bounded by $\tilde T_2$, as shown in
Eq.~\eqref{bell4p-T2}.  The signs are determined by the product
$\Xlami \Ylami$, which according to Table~\ref{tab:general1a} is
equal to 1 for $i=1,2,3$, and -1 for $i=4$.  Thus, if all terms
are to be positive, we need
\begin{equation}
\pxyli - \pxypli \,\, \begin{cases}
   \ge 0 & \hbox{for } i=1,2,3 \\
   \le 0 & \hbox{for } i=4 \ . 
   \end{cases}
\label{signs1}
\end{equation}

Next, we examine the fourth line of Eq.~\eqref{bell4p-E2},
which shows the terms that are bounded by $\tilde T_3$, as shown in
Eq.~\eqref{bell4p-T3}.  In this case the signs are controlled by 
the product $\Xlampi \Ylami$, which according to
Table~\ref{tab:general1a} is equal to 1 for $i=1,3,4$, and -1 for
$i=2$.  Thus, we require
\begin{equation}
\pxpyli - \pxpypli \,\, \begin{cases}
   \ge 0 & \hbox{for } i=1,3,4 \\
   \le 0 & \hbox{for } i=2 \ . 
   \end{cases}
\label{signs2}
\end{equation}

Finally, we examine the first two lines of Eq.~\eqref{bell4p-E2},
which are the terms that are bounded by $\tilde T_1$.  Arranging
the terms as in Eq.~\eqref{bell4p-T1s1}, the relevant signs are
determined by $\Ylami \Xlami$, which is 1 for $i=1,2,3$ and -1
for $i=4$; by $\Xlampi/\Xlami$, which is 1 for $i=1,3$, and -1
for $i=2,4$; and by $\Ylampi \Xlami$, which is 1 for $i=1,2,4$,
and -1 for $i=3$.  Using these signs, one can write the sum as
\begin{equation}
\begin{aligned}
\tilde T_1 &\le \Big| \Big\{ \big[ p(\lambda_1 | \x, \yp ) +
     p(\lambda_1 | \xp,\yp) \big]\\
  & \qquad \qquad + \big[ p(\lambda_1 | \x, \yp) - p(\lambda_1 |
     \xp, \yp ) \big] \\
& + \big[ p(\lambda_2 | \x, \yp ) + p(\lambda_2 | \xp,\yp) \big]\\
  & \qquad \qquad + \big[ p(\lambda_2 | \x, \yp) - p(\lambda_2 |
     \xp, \yp ) \big] \\
& + \big[ p(\lambda_3 | \x, \yp ) + p(\lambda_3 | \xp,\yp) \big]\\
  & \qquad \qquad + \big[ p(\lambda_3 | \xp, \yp) - p(\lambda_3 |
     \x, \yp ) \big] \\
& + \big[ p(\lambda_4 | \x, \yp ) + p(\lambda_4 | \xp,\yp) \big]\\
  & \qquad \qquad + \big[ p(\lambda_4 | \xp, \yp) - p(\lambda_4 |
     \x, \yp ) \big] \Big\} \Big| \ .
\end{aligned}
\end{equation}
Thus, the terms will all be positive provided that we insist that
\begin{equation}
\pxypli - \pxpypli \,\, \begin{cases}
   \ge 0 & \hbox{for } i=1,2 \\
   \le 0 & \hbox{for } i=3,4 \ . 
   \end{cases}
\label{signs3}
\end{equation}

To enforce these conditions, we parameterize the conditional
probability table as in Table~\ref{tab:f's}, which is designed so
that the inequalities of Eqs.~\eqref{signs1}, \eqref{signs2}, and
\eqref{signs3} are all enforced by the conditions $f_i \ge 0$ for
all $i$.

%%%%%%%%%%%%%%%%%%%%%%%%%%%%%%%%%%%%%%%%
%TABLE IX: f's
%%%%%%%%%%%%%%%%%%%%%%%%%%%%%%%%%%%%%%%%
\linespread{1.0} \renewcommand{\arraystretch}{1.0}
\begin{center}
\begin{table}[t]
\centering
\begin{tabular}{| c | c | c | c | c |} 
\toprule[1.5pt]
$\lam_i$        & $\pxyl$        & $\pxypl$ & $\pxpyl$ & $\pxpypl$ \\
\toprule[1.5pt]
${\lam}_1$ &
$f_1{\text +}f_5{\text +}f_6$ & $f_1{\text +}f_5$ & $f_1{\text +}f_7$ & $f_1$ \\
${\lam}_2$ &
$f_2{\text +}f_8{\text +}f_9{\text +}f_{10}$ & $f_2{\text +}f_8{\text +}f_9$ & $f_2$ & $f_2{\text +}f_8$ \\
${\lam}_3$ &
$f_3{\text +}f_{13}$ & $f_3$ & $f_3{\text +}f_{11}{\text +}f_{12}$ & $f_{3}{\text +}f_{11}$ \\
${\lam}_4$ &
$f_4$ & $f_{4}{\text +}f_{14}$ & $f_4{\text +}f_{14}{\text +}f_{15}{\text +}f_{16}$ & $f_4{\text +}f_{14}{\text +}f_{15}$ \\
\bottomrule[1.25pt]
\end{tabular}\par
\caption{Parameterization of the conditional probabilities $p
(\lambda_i \vert u,v)$, with the property that the inequalities
described by Eqs.~\eqref{signs1}, \eqref{signs2}, and
\eqref{signs3} are all enforced by the conditions $f_i \ge 0$ for
all $i$.}
\label{tab:f's}
\end{table}
\end{center}
%%%%%%%%%%%%%%%%%%%%%%%%%%%%%%%%%%%%%%%%

Since we would like our \four-parameter model to reduce to the
\two-parameter model of Sec.~\ref{sec:model} when $\Mtx = \Mx$ and $\Mty = \My$, it is useful to list the values of the $f$'s for the \two-parameter model (for $\My +
\Mtx + \Mty \le 2$):
\begin{equation}
\begin{aligned}
f_{1} &= (2 - \Mx - 2 \My)/8 \\
f_{2} &= (2 - \Mx - 2 \My)/8 \\
f_{3} &= (2 - \Mx - 2 \My)/8 \\
f_{4} &= (2 - \Mx - 2 \My)/8 \\
f_{5} &= (\Mx + \My)/4 \\
f_{6} &= 0 \\
f_{7} &= \My/2 \\
f_{8} &= \My/2 \\
f_{9} &= (\Mx - \My)/4 \\
f_{10} &= 0 \\
f_{11} &= (\Mx + \My)/4 \\
f_{12} &= 0, \\
f_{13} &= \My/2 \\
f_{14} &= \My/2 \\
f_{15} &= (\Mx - \My)/4 \\
f_{16} &= 0 \ .
\label{2pfs}
\end{aligned}
\end{equation}

The requirement that the model saturate the bound that $\ES \le 2 +
\My + \Mtx + \Mty$ can be expressed using Eq.~\eqref{bellE1} for
$\ES$, with Table~\ref{tab:general1a}.  The result can be written
most simply if one also uses the normalization of probabilities,
which gives
\begin{equation}
\begin{aligned}
S &= 4 - 2 \Big[ p(\lambda_1 | \xp,\yp) + p(\lambda_2 | \xp,\y) \\
  & \qquad +  p(\lambda_3 | \x,\yp) + p(\lambda_4 | \x,\y) \Big] \\
  &= 4 - 2 (f_1 + f_2 + f_3 + f_4 ) \, ,
\label{Sfromfs}
\end{aligned}
\end{equation}
so saturation implies that
\begin{equation}
f_1 + f_2 + f_3 + f_4 = 1 - \half (\My + \Mtx + \Mty) \, .
\label{saturation}
\end{equation}
Using this equation, the normalization equations are found to
be equivalent to
\begin{gather}
f_6 + f_{10} + f_{13} = f_{14} \, ,\label{fnorm1}\\
f_7 + f_{12} + f_{16} = f_8 \, ,\label{fnorm2}\\
f_5 + f_8 + f_9 + f_{14} = \half (\My + \Mtx + \Mty) \, ,\label{fnorm3}\\
f_8 + f_{11} + f_{14} + f_{15} = \half (\My + \Mtx + \Mty) \, .\label{fnorm4}
\end{gather}
We next calculate
\begin{align}
\Mx &= \Mx[\y] = f_{8} + f_{9} + f_{10} + f_{14} + f_{15} +
     f_{16} \nonumber \\
& \qquad + |f_{11} + f_{12} - f_{13}| + |f_{5} + f_{6} - f_{7}|
     \, ,\label{Mf1}\\
\My &= \My[x] =  f_{6} + f_{10} + f_{13} + f_{14} \, ,\label{Mf2}\\
\Mtx &= \Mx[\yp] =  f_{5} + f_{9} + f_{11} + f_{15} \, ,\label{Mf3}\\
\Mty &= \My[\xp] = f_{7} + f_{8} + f_{12} + f_{16}  \, .\label{Mf4}
\end{align}

By combining Eq.~\eqref{fnorm1} with Eq.~\eqref{Mf2}, and
Eq.~\eqref{fnorm2} with Eq.~\eqref{Mf4}, one has immediately
\begin{align}
f_{14} &= \half \My \, ,\label{f14}\\
f_{8} &= \half \Mty \, ,\label{f8}
\end{align}
and then Eqs.~\eqref{fnorm1}--\eqref{fnorm4} become
\begin{align}
f_6 + f_{10} + f_{13} &= \half \My \, ,\label{f6f10f13}\\
f_7 + f_{12} + f_{16} &=  \half \Mty \, ,\label{f7f2f16}\\
f_5 + f_9 &= \half \Mtx \, ,\label{f5f9}\\
f_{11} + f_{15} &= \half \Mtx \, ,\label{f11f15}
\end{align}
To make use of Eq.~\eqref{Mf1} for $\Mx$, one needs to evaluate
the two expressions inside absolute value signs.  From
Eq.~\eqref{2pfs}, we see that each expression is nonnegative in
the \two-parameter model.  Since we would like the \four-parameter
model to reduce to the \two-parameter model, we will assume that
these expressions are nonnegative here:
\begin{gather}
f_{11} + f_{12} - f_{13} \ge 0 \, ,\label{f11f12f13}\\
f_{5} + f_{6} - f_{7} \ge 0 \, ,\label{f5f6f7}
\end{gather}
in which case Eq.~\eqref{Mf1} simplifies to
\begin{equation}
f_7 + f_{13} = \half [ \My + \Mty - (\Mx - \Mtx)] \, .\label{f7f13}
\end{equation}

From Eqs.~\eqref{2pfs}, we see that for the \two-parameter
solution, $f_6 = f_{10} = f_{12} = f_{16} = 0$.  At this point we
will assume that the \four-parameter solution we seek maintains
the property that 
\begin{equation}
f_{10} = f_{16} = 0 \ ,
\label{f10f16}
\end{equation}
although we will see that it will not be possible to also require
$f_6$ and $f_{12}$ to vanish.  We will find such a solution,
which is our goal, and we make no claims that we will find all
solutions.  Then Eqs.~\eqref{f6f10f13} and
\eqref{f7f2f16} can be solved for $f_{13}$ and $f_7$, which
allows us to rewrite Eqs.~\eqref{f11f12f13}--\eqref{f7f13} as
\begin{align}
f_{11} &\ge \half (\My - \Mx + \Mtx) \, ,\label{f11ge}\\
f_5 &\ge \half (\Mty - \Mx + \Mtx) \, ,\label{f5ge}\\
f_6 + f_{12} &= \half(\Mx - \Mtx) \, ,\label{f6f12}
\end{align}
and the constraints $f_{13} \ge 0$ and $f_7 \ge 0$ become
\begin{align}
f_6 &\le \half \My \, ,\label{f6le}\\
f_{12} &\le \half \Mty \, .\label{f12le}
\end{align}

Consider first the values of $f_6$ and $f_{12}$. 
Eqs.~\eqref{f6f12}--\eqref{f12le} specify the sum of these two
quantities, and upper limits for each.  The limit for $f_6$ is
greater than or equal to the limit for $f_{12}$.  The sum may or
may not be smaller than the individual limits, but
Eq.~\eqref{const2} guarantees that the sum is always less than or
equal to the sum of the limits, so the equations can always be
satisfied.  A simple solution is to assign the full sum to
$f_6$, if the sum is less than the upper limit, and otherwise to
set $f_6$ equal to its upper limit, and assign the balance of the
sum to $f_{12}$:
\begin{align}
f_6 &= \half \min(\Mx{\text -}\Mtx,\My) \, ,\label{f6}\\
f_{12} &=  \half [\Mx - \Mtx - \min(\Mx{\text -}\Mtx,\My)]\, .\label{f12}
\end{align}
Given that we have chosen to set $f_{10}=f_{16}=0$,
Eqs.~\eqref{f6f10f13} and \eqref{f7f2f16} can now be used to show
that
\begin{align}
f_7 &= \half [\Mtx + \Mty - \Mx + \min(\Mx{\text -}\Mtx,\My)] \, ,\\
f_{13} &= \half [\My - \min(\Mx{\text -}\Mtx,\My)] \, .
\end{align}

Now consider the values of $f_5$ and $f_9$, where the sum is
given by Eq.~\eqref{f5f9} and a lower bound on $f_5$ is given by
Eq.~\eqref{f5ge}.  Both $f_5$ and $f_9$ must be nonnegative,
which may or may not be a more stringent bound for $f_5$ than
Eq.~\eqref{f5ge}, depending on parameters.  In addition,
Eq.~\eqref{2pfs} shows the values we would like these functions
to have when $\Mtx = \Mx$ and $\Mty = \My$.  A reasonably simple
solution satisfying all these properties is given by
\begin{align}
f_5 &= \frac{1}{4}[\Mtx + \My - \min(\Mx{\text -}\Mtx,\My)] \, ,\\
f_9 &= \frac{1}{4}[\Mtx - \My + \min(\Mx{\text -}\Mtx,\My)] \, .
\end{align}
The discussion of $f_{11}$ and $f_{15}$ is almost identical to
that of $f_5$ and $f_9$, except that the first terms on the
right-hand sides of Eqs.~\eqref{f11ge} and \eqref{f5ge} are
different.  But the same solution satisfies all the conditions:
\begin{align}
f_{11} &= \frac{1}{4}[\Mtx + \My - \min(\Mx{\text -}\Mtx,\My)] \, ,\\
f_{15} &= \frac{1}{4}[\Mtx - \My + \min(\Mx{\text -}\Mtx,\My)] \, .
\end{align}

Finally, we need to choose values of $f_1$--$f_4$ consistent with
the sum in Eq.~\eqref{saturation}.  Following the \two-parameter
expressions in Eq.~\eqref{2pfs}, we choose them to be equal, so
\begin{equation}
f_1 = f_2 = f_3 = f_4 = \frac{1}{8} (2 - \My - \Mtx - \Mty) \ .
\label{f1f2f3f4}
\end{equation}
All of the $f$'s have now been specified, and putting it all
together leads to Table~\ref{tab:4param1}.

To extend the model into the region $\My + \Mtx + \Mty > 2$, as
is shown in Table~\ref{tab:4param2}, there is again a systematic
method, but again it involves some arbitrary choices, so the
answer is not unique. 

Suppose that we are given an arbitrary allowed set of parameter
values, $(\Mx,\My,\Mtx,\Mty)$, consistent with
Eqs.~\eqref{const1} and \eqref{const2}, and the labeling
convention that $\Mx \ge \My$. Our goal is to construct a table
of conditional probabilities consistent with these parameters.

If $\My + \Mtx + \Mty \le 2$, then we of course just use the
solution already constructed.
But if the same table is used when $\My + \Mtx + \Mty > 2$, one
sees immediately that the terms on the diagonal running from
lower left to upper right (hereafter, the main diagonal) all
become negative.  Saturation for $\My + \Mtx + \Mty > 2$ implies
$S=4$, which with Eq.~\eqref{Sfromfs} implies that the sum of the
main diagonal terms must vanish, which in turn implies that each
term on the main diagonal must vanish, since they cannot be
negative. It is thus clear that for  $\My + \Mtx + \Mty > 2$, the
terms on the main diagonal of Table~\ref{tab:4param1} must be
adjusted by adding a quantity $2 h$, given by
\begin{equation}
h =  \frac{1}{2} \qthree = \frac{1}{16} (\My + \Mtx + \Mty - 2 ) \, ,
\end{equation}
where $\qthree$ is defined in Eq.~\eqref{qthree}.  We initially
allow arbitrary variation of the other entries, requiring however
that the sum for each row remain equal to 1.  Such an arbitrary
variation can be parameterized by the matrix $G_{i,j}$ shown in
Table~\ref{tab:g's}, where the full conditional probabilities for
$\My + \Mtx + \Mty > 2$ will be given by
\begin{equation}
P_{i,j} = P^{(0)}_{i,j} + G_{i,j} \ ,
\end{equation}
where $P^{(0)}_{i,j}$ is the matrix defined by
Table~\ref{tab:4param1}.

%%%%%%%%%%%%%%%%%%%%%%%%%%%%%%%%%%%%%%%%
%TABLE: g's
%%%%%%%%%%%%%%%%%%%%%%%%%%%%%%%%%%%%%%%%
\linespread{1.0} \renewcommand{\arraystretch}{1.0}
\begin{center}
\begin{table}[ht]
\centering
\begin{tabular}{| c | c | c | c | c |} 
\toprule[1.5pt]
$\lam_i$        & $\pxyl$        & $\pxypl$ & $\pxpyl$ & $\pxpypl$ \\
\toprule[1.5pt]
${\lam}_1$ &
$g_1+g_2$ & $-g_3 - h$ & $-g_5 - h$ & $2h$ \\
${\lam}_2$ &
$-g_1 - h$ & $g_3 + g_4$ & $2h$ & $-g_7-h$ \\
${\lam}_3$ &
$-g_2-h$ & $2h$ & $g_5 + g_6$ & $-g_8-h$ \\
${\lam}_4$ &
$2h$ & $-g_4-h$ & $-g_6-h$ & $g_7+g_8$ \\
\bottomrule[1.25pt]
\end{tabular}\par
\caption{Definition of the matrix $G_{i,j}$, where the matrix of conditional
probabilities for the four-parameter model, when $\My + \Mtx +
\Mty > 2$, is written as $P_{i,j} = P^{(0)}_{i,j} + G_{i,j}$,
where $P^{(0)}_{i,j}$ is the matrix in Table~\ref{tab:4param1}.}
\label{tab:g's}
\end{table}
\end{center}
%%%%%%%%%%%%%%%%%%%%%%%%%%%%%%%%%%%%%%%%

To prevent the calculations of $\Mx[v]$ and $\My[u]$ from
becoming prohibitively complicated, we will insist that the $g's$
be chosen so that the ordering of any two terms that are
subtracted in the calculations of $\Mx[v]$ and $\My[u]$ is
fixed.  Since we are trying to construct a four-parameter model
that reduces to the two-parameter model, we choose the ordering
to match that of the two-parameter model.  From
Table~\ref{tab:general2a}, we see that
\begin{equation}
\begin{aligned}
   & P_{1,1} - P_{1,2} \ge 0\, , && P_{1,3} - P_{1,4} \ge 0,\\
   & P_{2,1} - P_{2,2} \le 0\, , && P_{2,3} - P_{2,4} \le 0\, ,\\
   & P_{3,1} - P_{3,2} \ge 0\, , && P_{3,3} - P_{3,4} \ge 0\, ,\\
   & P_{4,1} - P_{4,2} \le 0\, , && P_{4,3} - P_{4,4} \le 0\, ,\\
   & P_{1,2} - P_{1,4} \ge 0\, , && P_{1,1} - P_{1,3} \ge 0\, ,\\
   & P_{2,2} - P_{2,4} \ge 0\, , && P_{2,1} - P_{2,3} \ge 0\, ,\\
   & P_{3,2} - P_{3,4} \le 0\, , && P_{3,1} - P_{3,3} \le 0\, ,\\
   & P_{4,2} - P_{4,4} \le 0\, , && P_{4,1} - P_{4,3} \le 0\, .
\label{orderings}
\end{aligned}
\end{equation}
Note that in two cases ($P_{2,1} - P_{2,2}$ and $P_{4,3} -
P_{4,4}$) these inequalities are inconsistent with
Eqs.~\eqref{signs1}, \eqref{signs2}, and \eqref{signs3}, but that
is expected.  Eqs.~\eqref{signs1}, \eqref{signs2}, and
\eqref{signs3} are the conditions to saturate $S \le 2 + \My +
\Mtx + \Mty$, but for $\My + \Mtx + \Mty > 2$, the bound to be
saturated is $S \le 4$.  For these two cases,
Table~\ref{tab:4param1} shows that, for $\My + \Mtx + \Mty \le 2$,
$P_{2,1} = P_{2,2}$ and $P_{4,3} = P_{4,4}$, so the orderings
specified in Eq.~\eqref{orderings} do not require any changes in
ordering as $\My + \Mtx + \Mty$ crosses the borderline at 2.

The other crucial requirement on the $g$'s is the positivity of
the conditional probabilities,
\begin{equation}
P_{i,j} \ge 0 \ .
\label{positivity}
\end{equation}

With the orderings specified in Eq.~\eqref{orderings}, it is
straightforward to find
\begin{equation}
\begin{aligned}
\Mx[\y] &= \Mx + 2 (g_2 + g_5 - 2 h) \, ,\\
\My[\x] &= \My + 2 (g_1 + g_3 - 2 h) \, ,\\
\Mx[\yp] &= \Mtx + 2 (g_4 + g_7 - 2 h) \, ,\\
\My[\xp] &= \Mty + 2 (g_6 + g_8 - 2 h) \, .
\end{aligned}
\end{equation}
A successful model requires that the second term on the
right-hand side of each line should vanish, which allows us to
solve for $g_5$, $g_3$, $g_7$, and $g_8$ in terms of the other
$g$'s.

The problem now is to find values for the independent $g_i$'s ---
$g_1$, $g_2$, $g_4$, and $g_6$ --- which are consistent with all
the constraints in Eqs.~\eqref{orderings} and \eqref{positivity}.

When the 32 constraints are written out, one finds that each of
the four independent $g_i$'s appears in 8 of them, with 4 in the
form of upper limits, and 4 in the form of lower limits.  In
every case there is one redundant pair, so each independent $g_i$
has three upper bounds and three lower bounds.  One of the upper
bounds and one of the lower bounds involves a second independent
$g$, so we put those bounds aside for later consideration.  This
leaves two upper bounds and two lower bounds for each independent
$g_i$.  Depending on parameters, either one of the upper bounds
and either one of the lower bounds can be the most restrictive. 
One can then construct a function equal to the minimum of the two
upper bounds and a function equal to the maximum of the two lower
bounds, so now one has one upper bound and one lower bound for
each independent $g_i$.  It can then be shown that if these
bounds are all satisfied, then the inequalities that we put aside
--- those that involve more than one independent $g_i$ --- are
automatically satisfied.

By comparing the bounding functions for the different $g_i$'s,
one finds that there are some simple regularities.  $g_{1,\rmax}$
is for all parameters at least as stringent as $g_{6,\rmax}$, so we
can take $g_{1,\rmax}$ as the upper bound for both $g_1$ and $g_6$,
where
\begin{equation}
\begin{aligned}
g_{1,\rmax} &= \frac{1}{16} \Big[R + 4 (2 \Mtx + \Mty + 2) \\
   & \qquad - 8 \max(\My + \Mtx, 2) + 4 \qtwo \Big] \, ,
\end{aligned}
\end{equation}
where $R$ and $\qtwo$ were defined in Eqs.~\eqref{Rdefs} and
\eqref{qtwo}, respectively.  Similarly, we can take $g_{6,\rmin}$
as the lower bound for both $g_1$ and $g_6$, where
\begin{equation}
\begin{aligned}
g_{6,\rmin} &= \frac{1}{16} \Big[R + 4 (2 + \Mty) \\ 
     & \qquad - 8 \min(\Mty+\Mx,2) + 4 \qtwo) \Big]\, .
\end{aligned}
\end{equation}

Since $g_1$ and $g_6$ now have the same upper and lower bounds,
we can choose to satisfy these relations by setting them equal to each
other, and equal to the mean of the upper and lower bounds:
\begin{equation}
g_1 = g_6 = \frac{1}{2} \Big[ g_{1,\rmax} + g_{6,\rmin} \Big] \, .
\end{equation}

A similar analysis of $g_2$ and $g_4$ shows that they can also be
described by common bounds, with
\begin{align}
g_{2,\rmax} &= \frac{1}{16} \Big[-3 R + 8 (\My - 2) \\
     &\qquad + 8 \min(\Mx+\Mty,2) - 8 \qtwo ) \, ,\\
g_{4,\rmin} &= \frac{1}{16} \Big[8 \max(\Mtx+\My,2) -3 R - 16
     \Big]\, .
\end{align}
We choose the solution
\begin{equation}
g_2 = g_4 = \frac{1}{2} \Big[ g_{2,\rmax} + g_{4,\rmin} \Big] \, .
\end{equation}

The final notation was chosen to simpify the appearance
of the solution, defining
\begin{align}
g_1 &= g_6 = h - q_4 \, ,\\
g_2 &= g_4 = h + q_4 \, ,
\end{align}
where $\qfour$ was defined in Eq.~\eqref{qfour}.

When the matrix $G_{i,j}$ is rewritten in terms of $q_3$ and
$q_4$, one finds the conditional probabilities given in
Table~\ref{tab:4param2}, thus completing the construction of the
four-parameter model.

\clearpage

%----------------------------------------------------------------------------------------------------------------------------------------------------------------------------------

%Don't use \bibliographystyle{} if you want longbibliography in \documentclass[12pt,aps,showpacs,notitlepage,longbibliography]{revtex4-1} to work where article titles are listed in the bibilography as well

%\bibliography{bib/selection,bib/sn2,bib/sngroup2,bib/jasonDK4,bib/jasonDK6,bib/mypapers2,bib/media2} 

%\bibliography{bib/oldbib/selection,bib/oldbib/sn2,bib/oldbib/sngroup2,bib/oldbib/jasonDK4,bib/oldbib/jasonDK6,bib/oldbib/mypapers2,bib/oldbib/media2} 

%\bibliography{bib/generalbound} 


\end{document}